\documentclass[a4paper,fleqn,usenatbib]{mnras}

\usepackage{graphicx}	
\usepackage{amsmath}	
\usepackage{amssymb}	

\newcommand\mj{{\,{\rm M}_{\rm J}}}
\newcommand\simlt{\la}

\newcommand\msun{M_\odot}
\newcommand\be{\begin{equation}}
\newcommand\ee{\end{equation}}
\newcommand\dr{\rm d}

\title[Migration code comparison]{Giant planets and brown dwarfs on wide orbits: a code comparison project.}

\author[Fletcher et al.]{
M. Fletcher$^{1}$, S. Nayakshin$^{1}$, D. Stamatellos$^{2}$, W. Dehnen$^{1,3}$, F. Meru$^{4,5,6}$,
L. Mayer$^{7}$, \newauthor
H. Deng$^{7}$, K. Rice$^{8,9}$\\
$^{1}$Department of Physics and Astronomy, University of
  Leicester, Leicester LE1 7RH, UK. {E-mail: mf240@le.ac.uk}\\
$^{2}$Jeremiah Horrocks Institute for Mathematics, Physics \& Astronomy, University of Central Lancashire, Preston, PR1 2HE, UK\\
$^{3}$Universit\"ats-Sternwarte der Ludwig-Maximilians-Universit\"at, Scheinerstra\ss e 1, M\"unchen D-81679, Germany\\
$^{4}$Department of Physics, University of Warwick, Gibbet Hill Road, Coventry, CV4 7AL, UK\\
$^{5}$Centre for Exoplanets and Habitability, University of Warwick, Gibbet Hill Road, Coventry CV4 7AL, UK\\
$^{6}$Institute of Astronomy, University of Cambridge, Madingley Road, Cambridge, CB3 0HA, UK\\
$^{7}$CTAC, Institute for Computational Science, University of Zurich, Winterthurerstrasse 190,
8057 Zurich, Switzerland \\
$^{8}$SUPA, Institute for Astronomy, Royal Observatory Edinburgh, University of Edinburgh, Blackford Hill, Edinburgh EH9 3HJ, UK \\
$^{9}$Centre for Exoplanet Science, University of Edinburgh, Edinburgh, UK
}

\date{Accepted XXX. Received YYY; in original form ZZZ}

\pubyear{2017}

\begin{document}
\label{firstpage}
\pagerange{\pageref{firstpage}--\pageref{lastpage}}
\maketitle

\begin{abstract}
Gas clumps formed within massive gravitationally unstable circumstellar discs are potential seeds of gas giant planets, brown dwarfs and companion stars. Simulations show that competition between three processes -- migration, gas accretion and tidal disruption -- establishes what grows from a given seed.  Here we investigate the robustness of numerical modelling of clump migration and accretion with the codes PHANTOM, GADGET, SPHINX, SEREN, GIZMO-MFM, SPHNG and FARGO. The test problem comprises a clump embedded in a massive disc at an initial separation of 120 AU. There is a general qualitative agreement between the codes, but the quantitative agreement { in the planet migration rate ranges from $\sim 10$\% to $\sim 50$\%}, depending on the numerical setup. We find that the artificial viscosity treatment { and the sink particle prescription} may account for much of the differences between the codes. { In order to understand the wider implications of our work, we also attempt to reproduce the planet evolution tracks from our hydrodynamical simulations with prescriptions from three previous population synthesis studies. We find that the disagreement amongst the population synthesis models is far greater than that between our hydrodynamical simulations. The results of our code comparison project are therefore encouraging in that uncertainties in the given problem are probably dominated by the physics not yet included in the codes rather than by how hydrodynamics is modelled in them.}
\end{abstract}

\begin{keywords}
Planet migration -- accretion discs -- hydrodynamics 
\end{keywords}

\section{Introduction}\label{sec:intro}
Secondary star formation via gravitational instability (GI) of massive circumstellar discs has now been observed by ALMA \citep{TobinEtal16}  { and may be a viable explanation for} the high frequency and the host metallicity correlations of stellar  binaries  with separations less than tens of AU \citep{MoeEtal18}. Modern star formation simulations \citep{Bate18} and observations of young discs \citep{TychoniecEtal18} also indicate that massive large gas discs { could} be abundant. 

The conditions for disc fragmentation \citep{Gammie01,Rafikov05} are similar to those for forming first hydrostatic cores in star formation \citep{Larson69}, implying that the masses of gas clumps born in the discs must be initially similar to those of the opacity-limited fragments, e.g., $\sim 5-10\mj$ \citep{Low76,Rees76,Masunaga98}, although both smaller and larger initial clump masses were considered in the literature \citep{BoleyEtal10,KratterEtal10,ForganRice13}. Due to these uncertainties and due to strong clump evolution after formation via inward migration \citep{MayerEtal04,VB05,MachidaEtal11,BaruteauEtal11}, gas accretion \citep{ZhuEtal12a,Stamatellos15,MercerStamatellos17} and tidal disruption \citep{BoleyEtal10,Nayakshin10c}, it is difficult to predict when and how often disc fragmentation leads to the formation of planets \citep{Kuiper51b}, brown dwarfs \citep{SW08,SW09} or secondary stellar companions \citep{KratterEtal10}.

The scale of uncertainty in this problem is immense and affects our understanding of even the most basic questions, especially in the theory of planet formation. Direct imaging surveys show that the occurrence rate of
wide separation (tens of AU or more) planetary mass companions to FGK stars, and also brown dwarfs, is just a few \% \citep{BillerEtal13,ChauvinEtal15,ReggianiEtal16,ViganEtal17}. This is much smaller than $\gtrsim 50$\% observed planet occurrence rate at separations less than a fraction of AU from the star \citep[see chapter 2 in][]{WF14}. One interpretation of this result is that gravitational disc instability rarely makes planetary-mass objects \citep{KratterEtal10,ForganRice13,RiceEtal15,ViganEtal17}. On the other hand, if radial migration and tidal disruption transmogrify planetary mass gas clumps into short period planets, including sub-Neptune mass planets \citep{BoleyEtal10,NayakshinFletcher15}, then 
the rate at which GI fragmentation forms planetary-mass clumps could be much higher; the resulting planets are simply not where they were born.

Furthermore, there is now observational support that at least some initially widely separated objects end up at sub-AU separations from the star, presumably due to disc migration. The frequency of appearance of planets more massive than $\sim 4$ Jupiter masses and brown dwarf companions to stars do not correlate with the host star metallicity \citep{RaghavanEtal10,TroupEtal16,Nayakshin_Review,SantosEtal17}, indicating that these objects probably did not form by Core Accretion \citep[which predicts an opposite correlation, see][]{MordasiniEtal12}. Additionally, the properties and statistics of very strong episodic flaring of young protostars, known as FU Ori outbursts \citep{HK96}, are consistent with stars tidally disrupting and devouring \citep{VB06,TakamiEtal18} up to a dozen gas clumps per lifetime.

There are many physical uncertainties in the physics of the problem, e.g., disc opacity \citep{MeruBate10b}, initial conditions for disc fragmentation \citep{VB10,ZhuEtal12a}, treatment of gas cooling close to and inside the Hill sphere of the planet \citep{NayakshinCha13,Stamatellos15,MercerStamatellos17}, dust growth and dynamics inside the clump, which may stronly affect clump cooling and heating balance \citep{HB11,Nayakshin16a}, etc.

However, in addition to this, different simulation codes use different numerical algorithms to model the same processes, and it is not clear if applying these codes to the same problem will yield identical results. 
The goal of our paper is evaluate how the simulation results differ between some commonly used numerical codes.  To focus on this issue alone, we set up a physically simple test problem of a gas giant planet embedded in a massive gas disc at an initial separation of 120 AU. 
The disc cooling is treated with the widely used idealised $\beta$-cooling { prescription} \citep{Gammie01,Rice05}. 

To disentangle various effects, we perform four comparison runs. The initial planet mass is set to $M_{\rm p0} = 2\mj$ in three of the runs and to $M_{\rm p0} =12\mj$ in the fourth. As explained above, gas accretion onto the gas clumps is an integral part of the problem. Therefore, in two of the $M_{\rm p0} = 2\mj$ runs we turn off gas accretion onto the planet, setting instead a relatively large gravitational softening length parameter to reduce the amount of gas flowing into the gravitational potential well of the planet \citep[as was also done by][]{BaruteauEtal11}. In the other two comparison runs, a sink particle prescription is used to absorb the gas accumulating at the planet location.

{ The paper is structured as follows. In \S \ref{sec:num} we describe the physical setup and initial conditions of the problem, and describe the contributing codes. In \S \ref{sec:results} we present main results of our paper. A comparison of the results to population synthesis prescriptions is made in \S \ref{sec:fits}, and in \S \ref{sec:discussion} we discuss observational implications of this work.}

%
%
\section{Problem and Numerical detail}\label{sec:num}

\subsection{Contributing Codes}\label{sec:codes}

There are five 3D SPH codes that we compare here: PHANTOM \citep{PriceEtal17}, GADGET \citep{Springel05}, SPHINX \citep{DehnenAly12}, SEREN \citep{HubberEtal11a, HubberEtal11b}, and SPHNG \citep[][]{Benz90}. The Meshless Finite Mass code GIZMO \citep{Hopkins15} builds on SPH methods and adds  a kernel discretization of the volume, coupled to a high-order matrix gradient estimator. The GIZMO-MFM numerical scheme has a higher order consistency and appears to overcome some of the numerical viscosity issues in SPH, and has been recently shown to reproduce the expected convergence of the critical cooling timescale for fragmentation \citep[see][]{DengEtal17}, which has been hard to achieve with SPH methods previously \citep[e.g.,][]{MeruBate10b}. Finally, FARGO is a 2D fixed cylindrical grid finite differencing code \citep{Masset00} which has been widely used for studies of planet migration and has shown consistency with analytical solutions in the linear regime applicable to much lower mass planets \citep[e.g.,][]{BaruteauMasset08} than studied here. 

\subsection{Problem choice}\label{sec:problem}

The potential formation of gas giant planets via gravitational instability of protoplanetary discs \citep[e.g.,][]{KratterL16} motivates our study.  To this end, all of our runs use a massive gas disc with initial mass $M_{\rm init} = 0.2\msun$  as an initial condition for all of our runs. The disc is in circular rotation around a star with mass $M_* = 1\msun$. At fragmentation, the disc \cite{Toomre64} $Q$-parameter is $Q\simlt 1$ \citep[e.g.,][]{BoleyEtal10}. Such discs  generate spiral density arms. Interactions of the planets with the arms give stochastic velocity kicks to the planets \citep[e.g.,][]{BaruteauEtal11}. In addition, fragmenting discs usually hatch more than one gas clump. Clump-clump interactions  also lead to angular momentum exchange between the clumps \citep{ChaNayakshin11a} and even mergers \citep{HallCEtal17}. These processes are stochastic and make numerical simulations of planet migration with different codes susceptible to small numerical detail.

To avoid this stochasticity, we simplified the task at hand by choosing the parameters of the problem such that  the Toomre parameter of the disc is slightly larger than expected at fragmentation, i.e., $Q \gtrsim 2$ everywhere, which makes the disc gravitationally stable. We then inject a planet into the disc and follow its evolution numerically. It is clearly desirable to extend the code comparison in the future in the regime in which the disc is free to fragment and form more clumps.


An ideal gas equation of state is used in this paper with the adiabatic index $\gamma = 7/5$, as appropriate for diatomic gas. The star irradiates the disc and sets the minimum irradiation temperature, which is a function of radius $R$:
\begin{equation}
T_{\rm irr} = T_0 \left(\frac{R}{R_0}\right)^{-1/2}\;,
\label{Tirr}
\end{equation}
where $T_0 = 20$~K and $R_0 = 100$~au. The irradiation temperature corresponds to the specific internal energy,
\begin{equation}
u_{\rm irr} = \frac{k_B T_{\rm irr}}{ \mu (\gamma-1)}\;,
\end{equation}
where $\mu = 2.45 m_p$ is the mean molecular weight of the gas.

The radiative cooling of the disc is modelled with the  $\beta$-cooling prescription widely used in the literature to model marginally stable self-gravitating discs \citep[e.g.,][]{Rice05}. The irradiation from the central star is additionally present as a heating term, so that the specific internal energy of the gas, $u$, evolves according to
\be
\frac{\dr u }{\dr t} = - \frac{ u - u_{\rm irr}}{t_{\rm cool}}\;,
\ee
where $t_{\rm cool} = \beta \Omega_{\rm K}(R)^{-1}$, 
\be
\Omega_{\rm K}(R) = \left( \frac{G M_*}{R^3} \right)^{1/2}
\label{eq:Kep}
\ee
We use $\beta=10$ for the runs presented below. { This value of $\beta$ is comparable to the critical fragmentation $\beta$ for $\gamma=7/5$ as found by \cite{Rice05}, although recent simulations with GIZMO-MFM suggest that disc fragmentation may occur at lower $\beta$ \citep[e.g.,][]{DengEtal17} for this code. However, the inclusion of external irradiation will also likely lead to fragmentation happening for lower values of $\beta$ (Rice et al., 2011)}

\subsection{Initial conditions}\label{sec:IC}

{ We first describe the initial conditions for the SPH codes and GIZMO-MFM.}
The star is treated as a sink particle that accretes any SPH particles that enter inside the sink radius, $R_{\rm sink} = 3$~au. The gravitational softening of the star is set at $h_{\rm g} = 0.25$~au. The disc is initially set up with the surface density profile
\be
\Sigma_{\rm in}(R) = {\frac{M_{\rm d}}{2\pi R (R_{\rm out} - R_{\rm in})}}
\ee
where $R_{\rm in} = 10$~au and $R_{\rm out} = 300$~au are the inner and the outer initial disc radii, respectively. The disc is relaxed for about 10 orbits at the outer edge before the planet is inserted. This is done to allow the disc to settle into a vertical hydrostatic balance and to damp out radial disc oscillations. During the disc relaxation procedure, a small fraction ($\sim 3$\%) of the SPH particles are accreted onto the central star. This is inevitable due to artificial viscosity of the disc increasing in regions of lower particle number, which is usually near the inner disc boundary.

\begin{figure} 
\includegraphics[width=0.99\columnwidth]{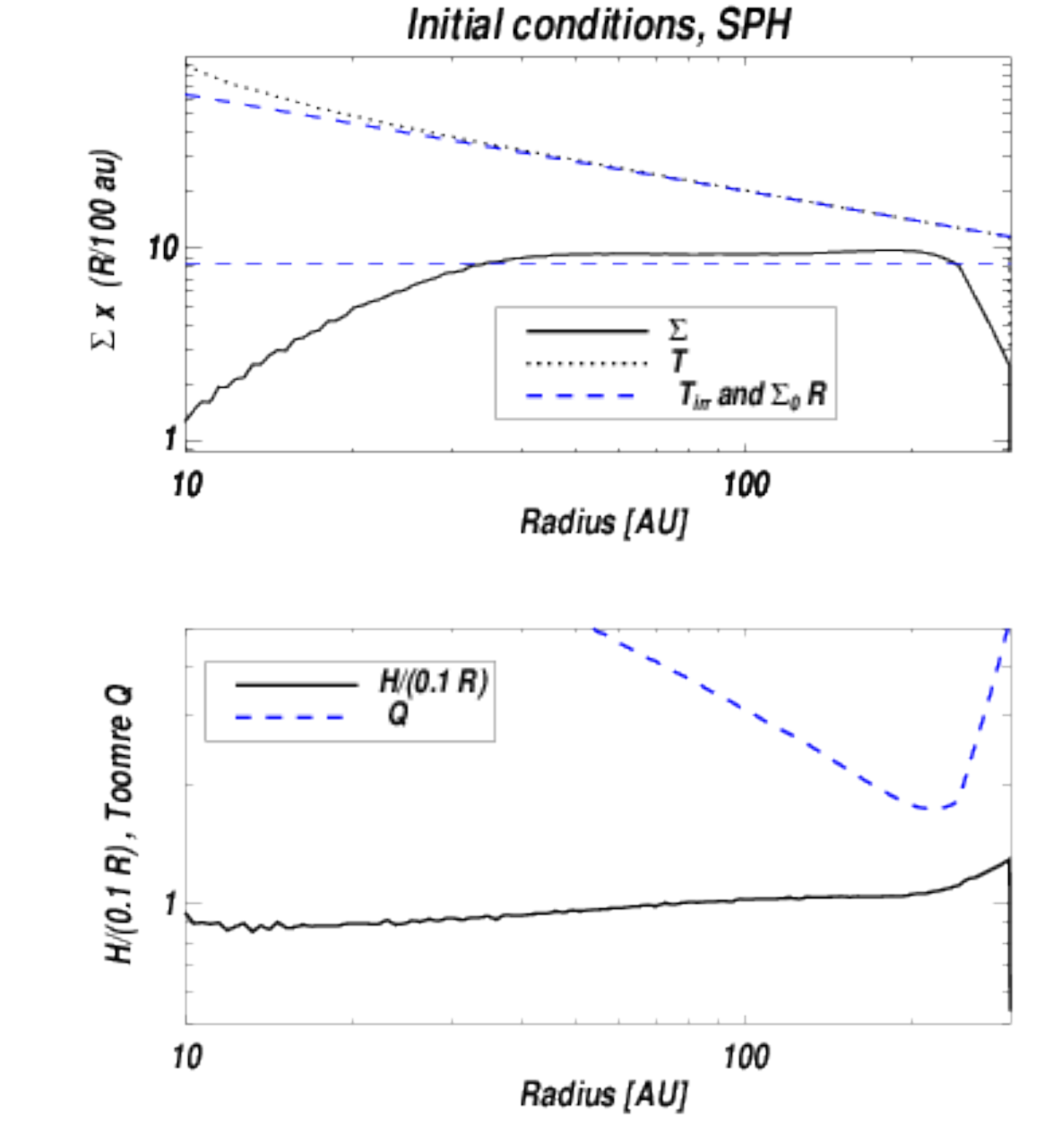}
\caption[Relaxed initial conditions for all of the SPH runs]{Initial (relaxed) conditions for all of the SPH runs presented in the paper. {\bf Top:} disc surface density, plotted as $\Sigma \times (R/100$~AU), and the temperature profiles. The disc inward of $\sim 30$~AU is strongly affected by the sink (star) particle inner boundary condition. {\bf Bottom:} The ratio of the disc vertical scaleheight $H$ to $0.1 R$ (solid) and the Toomre parameter $Q$. }
 \label{fig:disc_init}
 \end{figure}

These initial conditions, after the relaxation procedure was applied, are presented in fig. \ref{fig:disc_init}. The top panel shows the gas column density multiplied by radius, e.g., $\Sigma(R) \times (R/100$~AU)  and the vertically averaged gas temperature profile $T(R)$. Both of these are compared to the respective column density and temperature profiles before the relaxation (blue dashed curves). We see that both the inner and the outer regions of the disc are depleted by the relaxation process, but that the region between $R\sim 30$~AU and $R\sim 200$~AU has a smooth $\Sigma \propto 1/R$ profile. The gas temperature profile is very close to equation \ref{Tirr}, except for radii $R\lesssim 20$~AU where the artificial viscosity heating is not negligible. Since our relaxed disc has a strong roll-over at radii smaller than $R\sim 30$~AU, we expect that the planet migration process in this numerical setting will be strongly affected at radii of about $40$~AU. The bottom panel of Fig. \ref{fig:disc_init} presents the disc aspect ratio $H/R$ normalised to 0.1 and the Toomre $Q$-parameter. As stated in the Introduction, the disc is everywhere stable to self-gravity and does not fragment.

For FARGO, the initial conditions were obtained in the same physical setup but the code was relaxed for 50 orbits at the outer edge.

 In the simulations presented below, time is counted from the relaxed initial condition shown in Fig. 1. We inject the planet instantaneously at $t=0$ on a prograde circular orbit centred on the star at the initial separation of $R=120$~AU. No change is made to the initial velocity of either the gas or the star. Note that for $M_{\rm p0} = 2\mj$, the planet mass is only 1\% of the disc mass and just 0.2\% of the total mass of the system, so this approach is justifiable. While for the $M_{\rm p0} = 12\mj$ simulation the error is larger, we prefer this approach because keeping the planet orbit fixed while increasing its mass slowly (a common approach in studies of low mass planet migration) would lead to undesirable modifications of the disc structure for our problem. For example, as found by \cite{MalikEtal15}, the gap opening criterion should include a gap-opening time scale. If the planet migrates across the gap sooner than the gap could be excavated, no gap is opened. However, keeping the planet on a fixed orbit implies an infinite source of angular momentum and therefore may result in the planet opening a gap in the disc where none should be present.

\subsection{Approach to code algorithm differences}
 
Numerical hydrodynamics codes, whether particle or grid based, employ different numerical algorithms to integrate equations of motions, various time-stepping criteria, and approximate techniques to resolve contact discontinuities such as shocks and singularities arising in the gravitational potential and forces near point masses \citep{BodenheimerBook}. For example, by default GADGET uses the \cite{Monaghan97} form of the artificial viscosity with the Balsara-Monaghan switch to reduce artificial viscosity in shear flows \citep{Monaghan92,Balsara95}, and the spline kernel for SPH \citep[for details see][]{Springel05}. More modern formulations of artificial viscosity { exist} and different SPH kernels are adopted by some of the other codes (see \S \ref{sec:av}). It is possible to modify GADGET to use the same approaches. However, it is not possible in practice to modify all of the codes to employ exactly the same numerical algorithms due to significantly different intrinsic code designs. Additionally, such code alterations would defeat the purpose of our code comparison project as the codes actually being compared would then be different from their current community-used versions. 


Therefore, we attempted no code modification in this project with only a few exceptions that relate to the most salient physics of the problem. For each test problem presented below, all of the codes use the same gravitational softening parameters and the accretion radii  for the two sinks in the problem, as detailed further below. { The sections below discuss the implementation of sink particle accretion, gravitational softening and artificial viscosity in the codes used in this paper.}

\subsubsection{GADGET}\label{sec:gadget}

Our implementation of GADGET is very similar to the code description given in the instrument paper by \cite{Springel05}, with a few changes detailed below. GADGET uses the spline kernel \citep{MonaghanLattanzio85} for both the SPH density field and computing the gravitational softening around all particles, including the sink particles. We use 40 particles for the neighbour search. The artificial viscosity of SPH is that given by the Monaghan-Balsara formulation \citep{GingoldMonaghan82,Balsara95}, modified by the viscosity limiter prescription \citep[see eq. 11 in][]{Springel05}  to alleviate unwanted angular momentum transport in the presence of shear flows. 
We follow the default GADGET settings in this paper, keeping the artificial viscosity coefficient $\alpha_{\rm v}$ set to 1 for all times, and $\beta_{\rm v} = 2\alpha_{\rm v}$.

The sink particles are implemented in a very simple way. Any SPH particle that is separated from the sink by a distance smaller than the accretion radius $R_{\rm a}$ is accreted by the sink. The linear momentum and mass of the accreted particle are added to that of the sink. Some authors consider more complicated gas accretion criteria. For example, \cite{Bate95} consider the expected pressure of the gas within the sink region and the binding energy of the gas with respect to the sink. However, there is much physical uncertainty in picking these additional gas accretion criteria. The sink radius defines the region of space where we have insufficient information (usually, no information at all) about the gas properties. The interactions of that missing gas with the SPH particle in question could change the properties of the latter in ways that cannot be computed. For example, an SPH particle on a hyperbolic trajectory around the sink is formally not bound to the sink and thus would not be accreted if one accretes only particles with negative binding energies \citep{Bate95}. However, the same particle may be accreted if the particle were to interact with the missing gas within the sink radius, shock due to this interaction, and then lose the excess energy through radiation. 

For further discussion of these issues and tests of our GADGET implementation of the sink particle prescription, see \cite{Cuadra06} and \cite{HN18}. \cite{Nayakshin17a} found that the sink radius prescription tends to over-estimate the gas accretion rate onto a planet embedded in a massive gas disc for simulation parameters comparable to those used here \citep[see Fig. A1 in][]{Nayakshin17a}. Gas accretion rates measured in this paper should be thus taken as upper limits to the corresponding astrophysical problem.

\subsubsection{PHANTOM}\label{sec:phantom}
%
%

 \cite{CullenDehnen10} introduced an artificial viscosity switch which utilizes the derivative of the velocity divergence to detect shocks. Due to the switch, the artificial viscosity coefficient $\alpha_{\rm v}$ is varied between a minimum value, $\alpha_{\rm min}$, far from the shock, and the maximum, $\alpha_{\rm max} = 1$, reached close to the shock. We use this method for PHANTOM in this paper, as described in detail in \S\S 2.2.7-2.2.9 in \cite{PriceEtal17}. We fix the artificial viscosity coefficient $\beta_{\rm v}$ at 4 for our comparisons runs \citep[see][]{PriceFederrath10}. An exception to this is \S \ref{sec:av} where we explore how results depend on the choices of the artificial viscosity prescription for PHANTOM.

Gravitational softening in PHANTOM is different for interactions between sinks and interactions between sinks and SPH particles. The sink-sink softening is set to 0 by default. The sink-gas gravitational softening length is the maximum between the fixed softening length of the sink and the gas particles adaptive softening length. Gravity for SPH particles is softened by the SPH kernel function \citep[see \S 2.12.2 in][] {PriceEtal17}. 

Compared to GADGET, the PHANTOM default sink particle implementation also sets constraints on the binding energy and relative angular momentum of the SPH particle to be accreted. In this paper we disable these additional checks and use the same approach as specified in \S \ref{sec:gadget}.

\subsubsection{SPHINX}

SPHINX is an SPH code based on a conservative formulation (as derived from a variational principle, e.g.\ \citealt{Price12}) with individual artificial dissipation strengths $\alpha_{\rm v}$ adapted using the \cite{CullenDehnen10} switch with $\beta_{\rm v}=2\alpha_{\rm v}$. The details of the artificial viscous force differ slightly (by an amount $O(h^2)$) from traditional implementations to accommodate the one-sweep SPH algorithm, which avoids separate sweeps over all particle neighbours for the density and force computations. For the runs here, we use the \cite{Wendland1995} $C^2$ smoothing kernel, which { scales as} $w\propto(h-r)^3(h+3r)$ for $r<h$ with smoothing length $h$, adjusted to obtain $N_h=4\pi\rho h^3/3m=80$ at each time step. Gravity is computed using a $C^\infty$ softening kernel with density $\propto(r^2+h_{\rm s}^2)^{-7/2}$, which results in a smaller force bias than traditional Plummer softening \citep{Dehnen2001}. Individual softening lengths $\epsilon$ are scaled to the smoothing lengths $h$ such that the estimates for the gas and gravitating mass densities are mutually consistent (have the same bias). SPHINX uses an oct-tree for neighbour search (and gas-selfgravity which is computed using the fast multipole method \citealt{Dehnen2000}) and the leap-frog (2nd order symplectic) time integrator. Star and planets are represented by sink particles, whose gravity is computed by direct summation. Any gas particle within one sink radius is accreted by a sink particle, whereby its mass, linear and angular momentum, as well as energy is absorbed by the sink particle (which carries a spin and internal energy for this book keeping).

\subsubsection{SPHNG}

SPHNG is based on the version developed by \citep{Benz90} and first presented by \citep{Bate95}.  It uses variable individual smoothing lengths $h_j$ and adjusts them so that the number of nearest neighbours for any particle is $50 \pm 20$.  It also uses individual particle time-steps to simulate dense regions with sufficient precision while avoiding over-simulation of less dense regions, and integrates the particles using a second order Runge-Kutta scheme.   The standard artificial viscosity \citep{Monaghan92}, with $\alpha_{\rm v} = 1.0$ and $\beta_{\rm v} = 2.0$, and standard spline kernel are used.  A binary tree is used to calculate neighbour lists and to determine gravitational forces between gas particles, with the gravitational force softened by the SPH kernel function \citep{PriceEtal17}.  The gravitational force between the gas particles and the sink particles is, however, done using a direct calculation, which { is} softened by replacing the $1/r^2$ gravitational force dependence with $1/(r^2 + h_{\rm s}^2)$.  If accretion onto the sink particles is allowed, then particles are only accreted if they are bound and if the specific angular momentum of the particle is less than that required for them to form a circular orbit at the accretion radius \citep{Bate95}.

\subsubsection{GIZMO}

The GIZMO code is a multi-method code which inherits the tree-based gravity algorithm from GADGET3 \citep[see][for GADGET2 code description]{Springel05} and couples it with
different Lagrangian hydrodynamical solvers. For this paper we employ the Meshless Finite Mass (MFM) hydro method in GIZMO which solves the inviscid
fluid  equations by partitioning the computational domain using volume elements associated with a particle distribution, and computing fluxes through the volume ‘overlap’ by means of a Riemann solver as in finite volume Godunov-type methods \citep{Hopkins15}. Volume elements are constructed via convolution integrals with kernel functions analogous to those adopted in SPH. Owing to the use of a Riemann solver (here we use the HLLC solver and the minmod slope limiter), GIZMO-MFM employs no explicit artificial viscosity. This numerical method appears significantly less dissipative than SPH for differentially rotating flows, better conserving  angular momentum and vorticity \citep{Hopkins15,DengEtal17}. The kernel
for the volume partitioning, the gravitational softening and the sink particle implementation are all identical to those of GADGET used in this paper (\S \ref{sec:gadget}).

\subsubsection{SEREN}

The SPH code {\sc seren} was developed for star and planet formation simulations by \cite{HubberEtal11b, HubberEtal11a}. The code  uses an octal tree to compute gravity and find neighbours, multiple particle timesteps, and a 2$^{\rm nd}$ order Runge-Kutta integration scheme. To simulate the effect of physical viscosity in discs, {\sc seren} uses a time-dependent artificial viscosity \citep{MorrisMonaghan97} with parameters $\alpha_{\min}=0.1$, $\alpha_{\max}=1$ and $\beta_{\rm v}=2\alpha_{\rm v}$,  so as to reduce artificial shear viscosity away from shocks \citep[this scheme is the predecessor of the][method]{CullenDehnen10}. Sink particles, which  interact with the rest of the computational domain  only through their gravity, are used to represent the central star and the planet  \citep{BateEtal95}. Gas particles accrete onto a  sink when they are within the sink radius and bound to the sink \citep[see][]{HubberEtal11a}. Once gas particles are accreted, their mass and linear angular momentum is added to sink. The gravitational force between gas particles and a sink is found through a direct calculation and softened according to $1/(r^2+h_{\rm s}^2)$ to avoid unphysically large gravity forces.

\subsubsection{FARGO}

 FARGO is a 2D grid based, staggered-mesh code \citep{Masset00,BaruteauMasset08} that has been used extensively to study planet migration \citep{Masset02,MassetCasoli10,BaruteauEtal11}. For the runs presented here, we use a cylindrical grid with 508 and 1536 cells in the radial and azimuthal directions, respectively. The radial grid is logarithmic with the inner and outer boundary conditions set at 10 and 300 AU, respectively. Von Neumann--Richtmyer artificial bulk viscosity is used to treat contact discontinuities \citep{StoneNorman92}.

For this paper, FARGO also uses a fixed gravitational softening parameter $h_s$ as for all the other codes, which is a break with the common practice of scaling $h_s$ with the local disc scaleheight or the star-planet separation \citep[e.g.,][]{BaruteauEtal11}, but allows for a more uniform comparison between the codes. Specifically, the softening parameter used in grid simulations is typically set to $h_s = \epsilon H$ where $\epsilon \sim O(0.1)$ \citep{MullerEtal12a}. In this case the gravitational softening would be a function of position as $H \propto R$ for our simulations (see fig. \ref{fig:disc_init}). The consequences of this for numerics are not immediately obvious, but we note that for $\epsilon = 0.1$ and $H \sim 0.1 R$, the adaptive softening is equivalent to $h_{\rm s} = 0.4 - 1.2$~AU in the radial range 40-120 AU, which is not too dissimilar from the 1 AU and 2 AU fixed smoothing employed in Runs 1 and 2 (see below). For a relatively large fixed value $\epsilon = 0.7$ we find that the FARGO migration timescales increase by $\approx 50\%$ compared to those presented in this paper.

\subsection{The comparison runs}\label{sec:runs}

It is possible to resolve the pre-collapse gas giant planets (clumps) in modern computer simulations directly \citep[e.g.,][]{BoleyEtal10,GalvagniEtal12,ZhuEtal12b,Nayakshin17a,HallCEtal17}. However, while the clumps can be resolved and modelled from the point of view of hydrodynamics, other physics, e.g., a proper equation of state including molecular hydrogen internal degrees of freedom, dust dynamics and radiative transfer, are not yet implemented in most of the codes available to us here. Any simplified radiative transfer scheme applied to the clumps would necessarily over-simplify their internal physics \citep[their cooling balance is significantly different from that of the disc; e.g., see][]{VazanHelled12} and would thus be riddled with its own uncertainties.
A more prudent approach for us to follow here is to model the planet as a sink particle, just as the star, albeit with its own gas accretion (sink) radius.

Table \ref{table1} shows the parameters that distinguish the four different comparison runs that are presented below. In Runs 1-3, the initial planet mass is set to $M_{\rm p0} = 2\mj$, whereas Run 4 starts with $M_{\rm p0} = 12\mj$. In Run 1 and 2, gas accretion onto the sink is completely turned off by setting the accretion radius to zero. This is done to try to isolate the effects of planet migration versus gas accretion onto the planet. This is especially important since FARGO is a grid based code in which implementation of gas accretion is drastically different from the sink particle method of SPH codes. Therefore, Runs 1 and 2 can be simulated with SPH codes and FARGO, whereas Runs 3 and 4 are done with SPH only.

Turning off gas accretion onto a planet does not come free of numerical cost. Gas that gets bound to the planet may eventually get very close to the planet. A very high gas density around the planet is numerically challenging as the SPH particle time step becomes too short for the code to execute effectively. Therefore, to avoid that, in Runs 1 and 2 the planet softening radius, $h_{\rm s}$ is increased to 1 and 2 AU, respectively, from the much smaller value used in Run 3. For a similar reason Run 4 uses a larger accretion radius than Run 3. 

The initial SPH particle number is $N = 10^6$ for all of the runs presented here.

\begin{table}	
	\begin{center}
    \caption{The parameters distinguishing the Runs presented in this paper. $R_{\rm a}$, $h_s$, and{ $M_{\rm p0}$ are the sink accretion radius, the gravitational softening parameter, and the mass of the planet, }respectively. All the other parameters and initial conditions are the same for all four Runs.}
	\begin{tabular}{ l | c | c | c }
	\hline 
    Run & $R_{\rm a}$ (AU) & $h_{\rm s}$ (AU) & $M_{\rm p0}$ ($\mj$)  \\
    \hline \hline
	Run 1 & 0.0 & 1 & 2 \\
    Run 2 & 0.0 & 2 & 2 \\
    Run 3 & 0.5 & 0.01 & 2 \\
    Run 4 & 1.0 & 0.01 & 12 \\
    \hline
	\end{tabular}
    \label{table1}
    \end{center}
\end{table}


%
%

\subsection{Analytical expectations}

\cite{Tanaka02} derived an analytical expression for type I migration of a low mass planet in an isothermal disc. The migration timescale, defined as
\begin{equation}
\tau = \frac{R}{|\dot R|}\;,
\label{eq:tau}
\end{equation}
where $\dot R$ is the rate of change of planet-star separation due to gravitational torques from the disc, is given by
\begin{equation}
\tau_{\rm iso} = \left(2.7 + 1.1 \lambda \right)^{-1} \frac{M_{\star}}{M_{p}} \frac{M_{\star}}{\Sigma_{\rm p}r^{2}_{p}} \left(\frac{c_{s}}{r_{p}\Omega_{p}} \right)^{2} \frac{1}{\Omega_{p}}\;.
\label{eq:Tanaka}
\end{equation}
Here $\lambda$ is the exponent of the surface density power law, $\Sigma \propto R^{-\lambda}$, $\Sigma_{\rm p}$ is the surface density at the planet location, $M_{\star}$ and $M_{p}$ are the star and planet masses, respectively, $r_{p}$ is the planet-star separation, $c_{s}$ is the gas sound speed at the planet and $\Omega_{p}$ is the planet Keplerian angular velocity. For the initial parameters of our disc and $M_{\rm p}=2\mj$, we obtain a migration time scale of $\tau_{\rm iso} = 14.6 \times 10^3$~yr. Even though our discs are not isothermal, the results of \cite{Tanaka02} are widely used, and serve as a useful comparison for us.

\cite{BaruteauEtal11} used the 2D code FARGO to study planet migration in very massive self-gravitating discs, for which the Toomre parameter $Q$ self-regulates to a value between $\sim 1.5$ and $\sim 3$ over a broad range of radii. These authors also offered an analytical expression for the migration time scale:
\begin{equation}
\tau_{\rm sg} \approx \; \frac{5.6}{ (3.8 - \lambda)} \; \gamma Q_{p} \;\frac{h_{p}^{3}}{q} \;\left(\frac{0.1}{h_{p}}\right)^{2} \frac{2\pi}{\Omega_{p}}\;,
\label{eq:Baruteau}
\end{equation}
where $q = M_{\rm p}/M_{\star}$ is the mass ratio; $Q_p$ is the Toomre parameter and $h_p = H/R$ at the planet position. For the initial parameters of our Runs 1-3, eq. \ref{eq:Baruteau} yields $\tau_{\rm sg} = 5.0 \times 10^3$ yr at a separation $R=120$~AU.

%
%

\section{Results}\label{sec:results}

\subsection{At a glance}\label{sec:at_glance}

\begin{figure*}
\includegraphics[width=0.99\columnwidth]{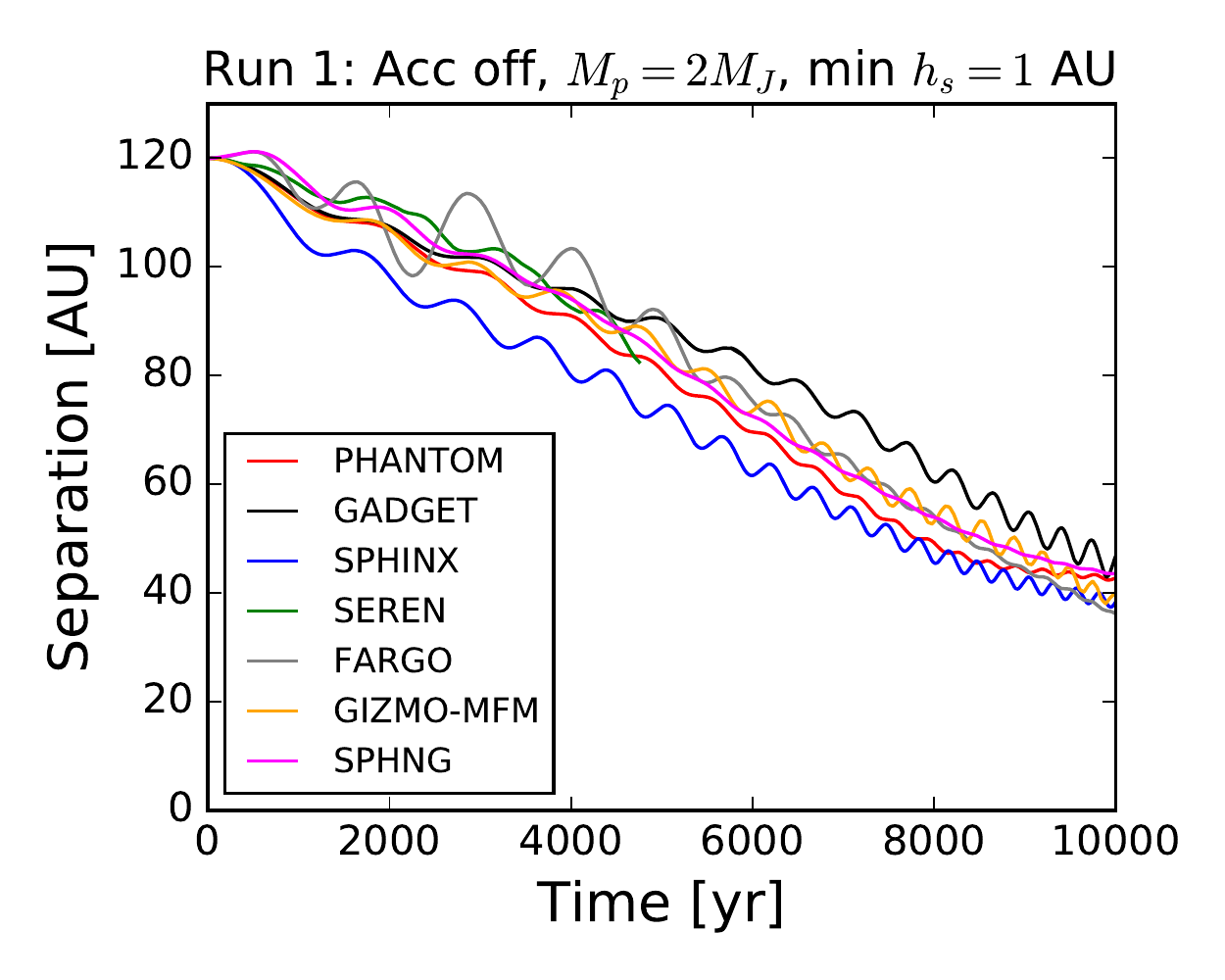}
\includegraphics[width=0.99\columnwidth]{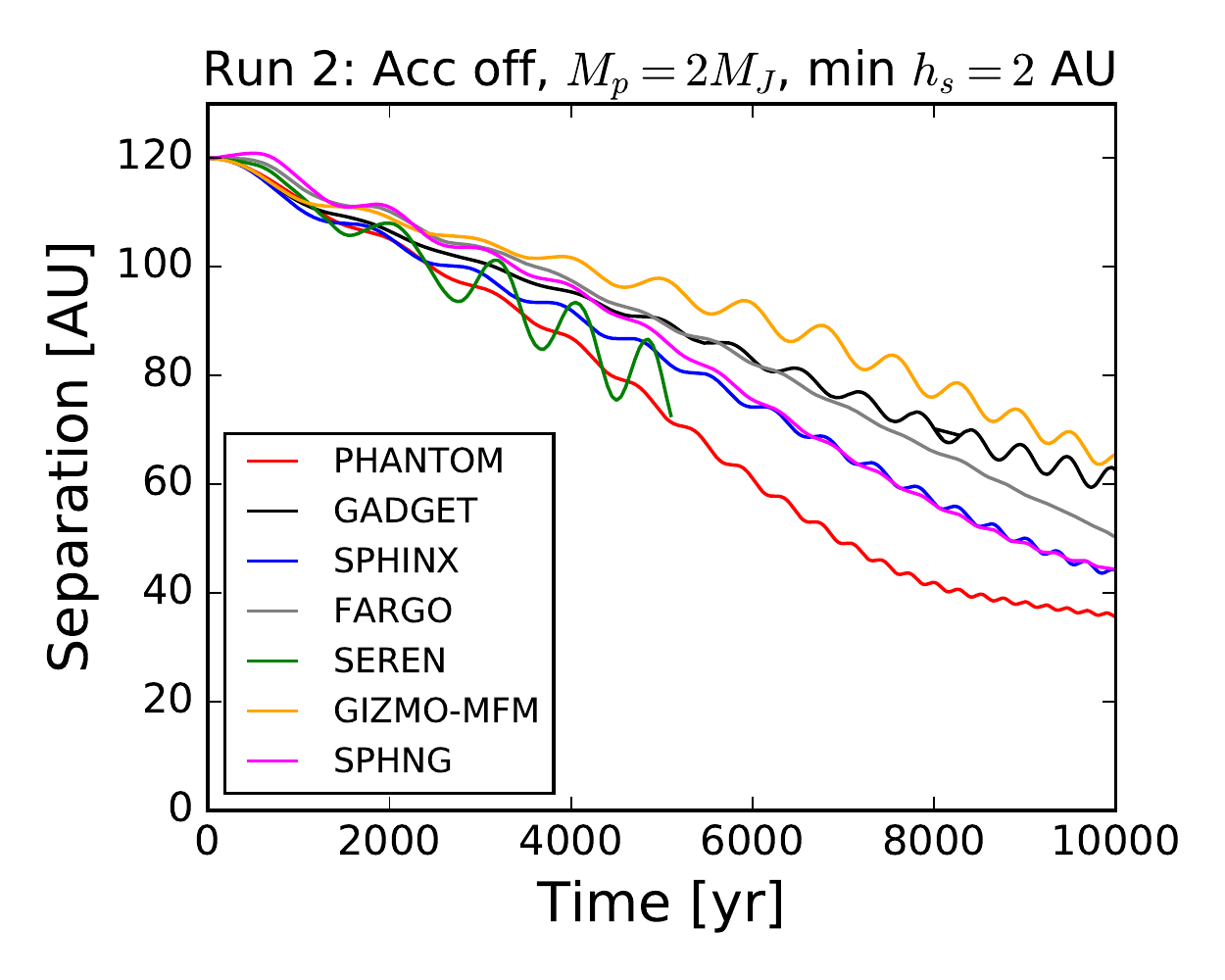}
\caption{Planet separation versus time for Run 1 (left panel) and Run 2 (right panel). Both of these do not allow the planet to gain mass from the disc, so the sink mass is fixed at $M_{\rm p0} = 2\mj$.}
\label{fig:run12}
\end{figure*}
\begin{figure*}
\includegraphics[width=0.99\columnwidth]{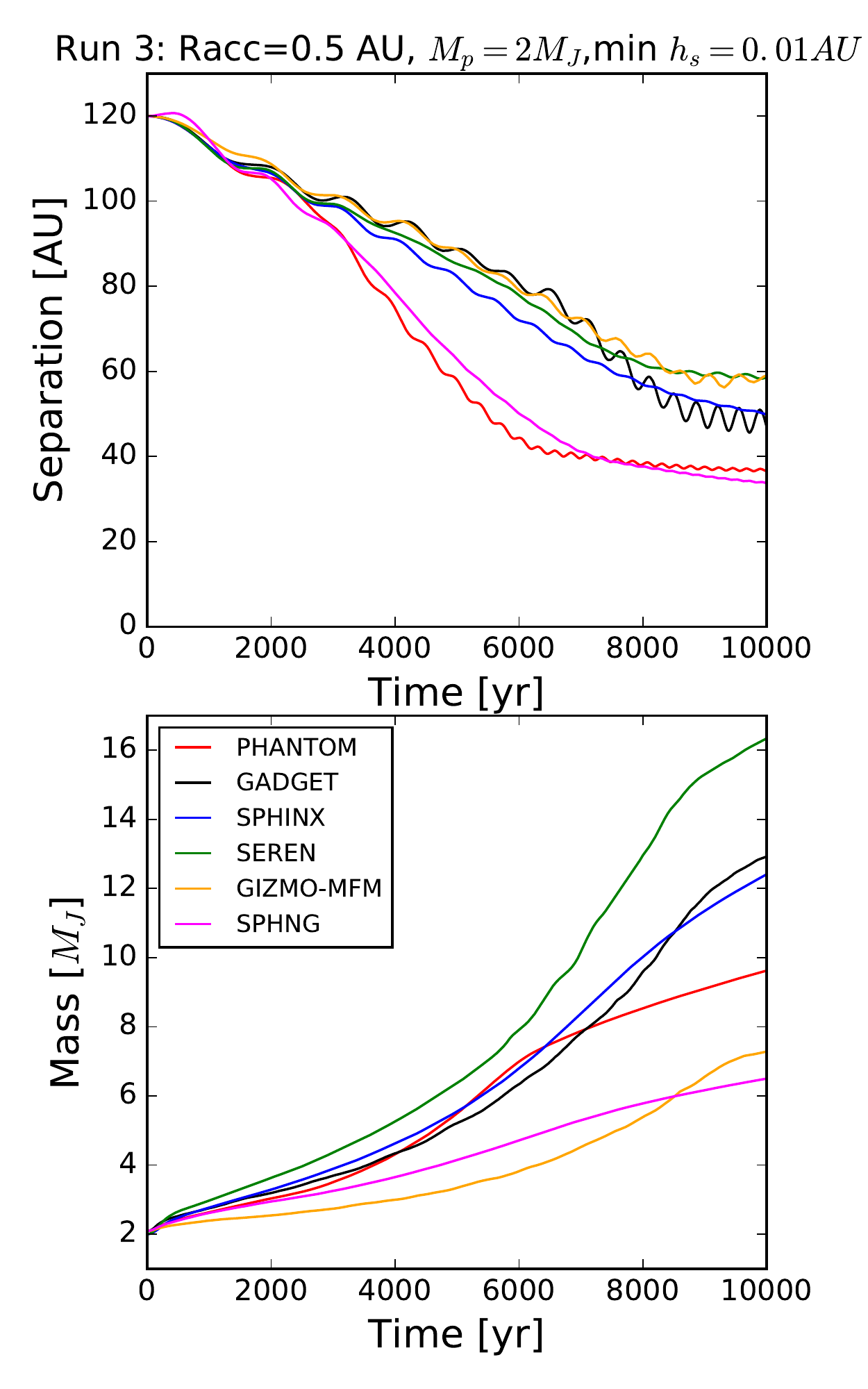}
\includegraphics[width=0.99\columnwidth]{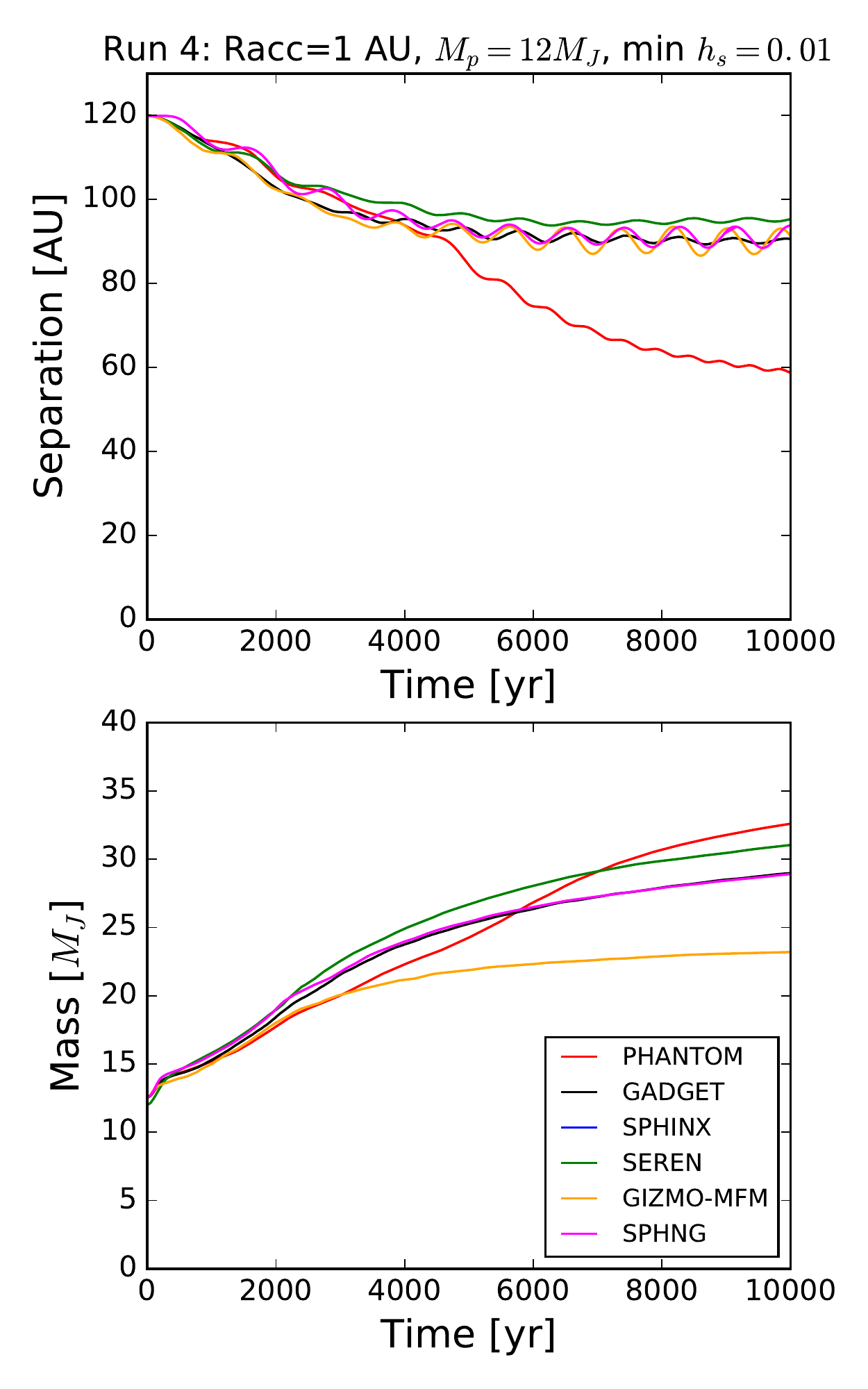}
\caption{Planet separation (top panel) and sink particle mass (lower panel) versus time for Runs 3 \& 4 (left and right panels, respectively).}
\label{fig:run34}
\end{figure*}

Fig. \ref{fig:run12}  shows the planet separation against time for Runs 1 \& 2. To recap, gas accretion onto the planet is off, and instead a relatively large gravitational softening parameter is used. Despite this, some gas accumulates deep inside the Hill sphere, and differently so for different codes. This appears to be the primary reason why Runs 1 and 2 stalled for SEREN at around 5000 yr. Fig. \ref{fig:run34} shows the results of Runs 3 \& 4 (left and right panels, respectively) in which gas accretion onto the planet (sink particle) is allowed. The sink mass versus time is shown in the lower panels. 

A cursory look at figs. \ref{fig:run12} \& \ref{fig:run34} shows that there is a general {\em qualitative} agreement between the different codes. For example, in Runs 1-3 the planet manages to migrate to separations of $40-60$~AU for most of the codes, whereas in Run 4, in which the planet is much more massive, the planet stalls further out due to it opening a deep gap in the disc. At the same time, there are significant {\em quantitative} disagreements between the codes. All of the codes show that the planet develops orbital eccentricity, but the actual value of the eccentricity is different, varying between $\sim 0.01$ to the maximum of $\sim 0.1$. 

\subsection{Analysis of Runs 1-3}\label{sec:run13}

\subsubsection{Migration rates}\label{sec:tmigr13}

We now analyse Runs 1-3 in which the planet initial mass is $M_{\rm p0} = 2\mj$. To aid quantitative analysis, we determine migration time scale, $\tau$, from the simulations. A straight-forward use of eq. \ref{eq:tau} to calculate $\tau$ from the simulation data is ill advised due to planets having non-zero eccentricity: the instantaneously defined migration time varies significantly over a fraction of the planet orbital timescale. Some sort of time averaging of $\tau$ over times at least as long as an orbital period is thus needed.

To do so, we first define a time-dependent migration rate as the final difference $\Delta R/\Delta t$, where the separation and time differences are counted from the initial values:
\begin{equation}
\dot R(t) = \frac{R(t) - R_0}{t}\;,
\label{dotR_sim}
\end{equation}
where $R_0 = R(t=0) = 120$~AU, $t>0$ is time, and $R(t)$ is the planet-star separation at that time. 
To remedy the oscillatory behaviour in $\dot R$ due to finite orbital eccentricity, we define an orbit-averaged quantity
\begin{equation}
\dot {\bar R}(t) = \frac{1}{T_{\rm p}} \int_{t-T_{\rm p}/2}^{t+T_{\rm p}/2} \dot R(t') {\rm d} t' \;,
\end{equation}
where $T_{\rm p}$ is the planet orbital period at location $R(t)$. We use this definition to define the planet migration rate after $t=4,000$~yr for all of the codes, which we label $\tau_4$. We then also define the migration time scale $\tau_7$, following the procedure outlined above, but uding the data between 4,000 and 7,000 yrs. Comparison of $\tau_4$ and  $\tau_7$ tells us how the migration rate varies as the planet gets closer to the star. Due to a non zero planet orbital eccentricity a finer time-resolved analysis of the migration rate does not appear well justified.


\begin{figure*}
\begin{center}
\includegraphics[width=0.99\textwidth]{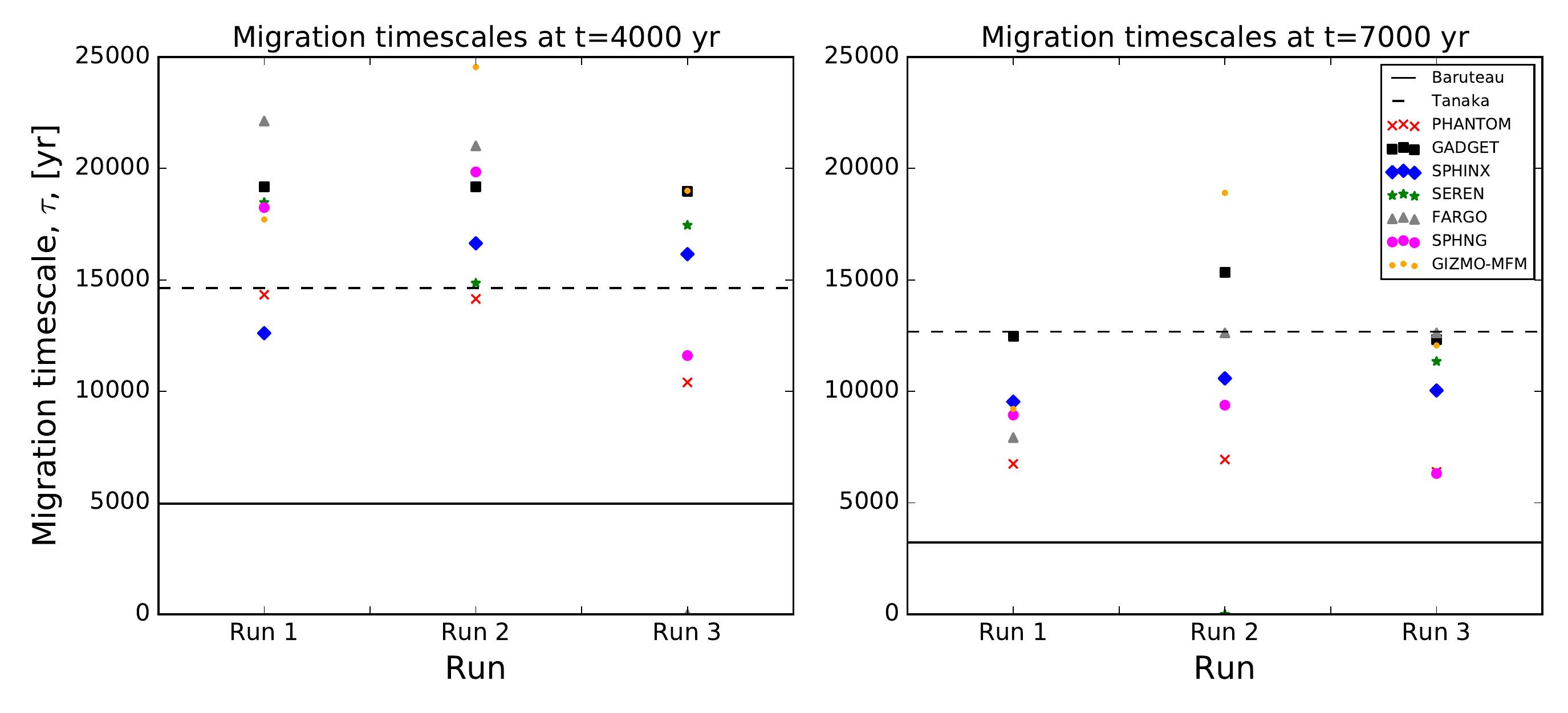}
\caption{Migration time scales for all codes for Runs 1-3 are shown with the coloured symbols, calculated for time intervals between 0 and 4,000 years {\bf (left panel)} and between 4,000 and 7,000 years {\bf (right panel)}. 
The dashed and solid horizontal lines show the analytically computed migration times given by eqs. \ref{eq:Tanaka} and \ref{eq:Baruteau}, respectively. The SEREN results do not appear for run 2 on the right panel since the code did not progressed to the 7000 year point.}
\label{fig:migration_timescales}
\end{center}
\end{figure*}

Fig. \ref{fig:migration_timescales} compares the migration time scales $\tau_4$ and $\tau_7$ (left and right panels, respectively) for all the codes for Runs 1-3, which are shown with the coloured symbols. The dashed and solid horizontal lines show the migration timescales given by eq. \ref{eq:Tanaka} and \ref{eq:Baruteau}, respectively. These analytical estimates of $\tau$ are computed using the initial disc properties (see fig. \ref{fig:disc_init}).

Taking the full range of $\tau_4$ and $\tau_7$ values, we see that they vary by a factor of $3-4$ between the different codes for Runs 1 \& 2, and by a smaller factor of $\sim 2$ for Run 3. For $\tau_4$, the mean of the migration time scales are closer to the \cite{Tanaka02} expression, but for $\tau_7$ the mean lies between the analytical estimates of \cite{Tanaka02} and \cite{BaruteauEtal11}. The range in the migration time scales is similar to the factor of $\sim 3$ difference between these two analytical results. We also note that $\tau_4$ is longer than $\tau_7$ for most of the runs, implying that migration of the planet accelerates somewhat as the planet gets closer to the star (as long as it remains in the Type I). The same trend is predicted by the formulae shown in equations \ref{eq:Tanaka} and \ref{eq:Baruteau}. We conclude from fig. \ref{fig:migration_timescales} that there is a qualitative agreement not only between the different codes but also with the theory.

Comparing Runs 1 and 2, we note that the migration timescales vary by $\sim 10$\% for most of the codes whereas $h_{\rm s}$ changes by a factor of two. However, for SEREN the difference { the two runs is larger, and is in the opposite sense compared with most of the other codes}. This is likely due to a non linear interplay of how gravitational softening affects gravitational torques vs planet accretion. To make further progress we must consider the role of planet accretion in greater detail, as the evolving planet mass certainly affects the migration rate.

\subsubsection{Hill mass versus sink mass}\label{sec:accr13}

The sink particle mass may not always properly reflect the mass of the planet. To quantify this, we define an effective Hill mass of the planet, $M_{\rm H}$, as the sink mass plus the mass of the gas within $R_{\rm H}/2$ of the planet. The choice of $R_{\rm H}/2$ is motivated by results of \cite{Nayakshin17a}, who finds that gas bound to the planet is usually located within half the Hill radius; material between $R_{\rm H}/2$ and $R_{\rm H}$ is much more likely to be lost as the planet migrates inwards. 

We should also note that the Hill radius definition needs to include the mass of the gas envelope around the sink itself, that is,
\begin{equation}
R_{\rm H} = R \left(\frac{M_{\rm H}}{3 M_{\star}} \right) ^{1/3}\;,
\label{eq:rhill}
\end{equation}
where we use $M_{\rm H}$ rather than the sink mass, $M_{\rm p}$. When the Hill mass is dominated by the sink mass, $M_{\rm H}$ can be safely replaced by $M_{\rm p}$, and the calculation of $M_{\rm H}$ from the particle data is trivial. In general, however, the mass of the gas surrounding the sink is not negligible, so we iterate over $R_{\rm H}$ and $M_{\rm H}$ to find self-consistent values for these two quantities that obey eq. \ref{eq:rhill}.

\begin{figure*}
\includegraphics[width=0.99\textwidth]{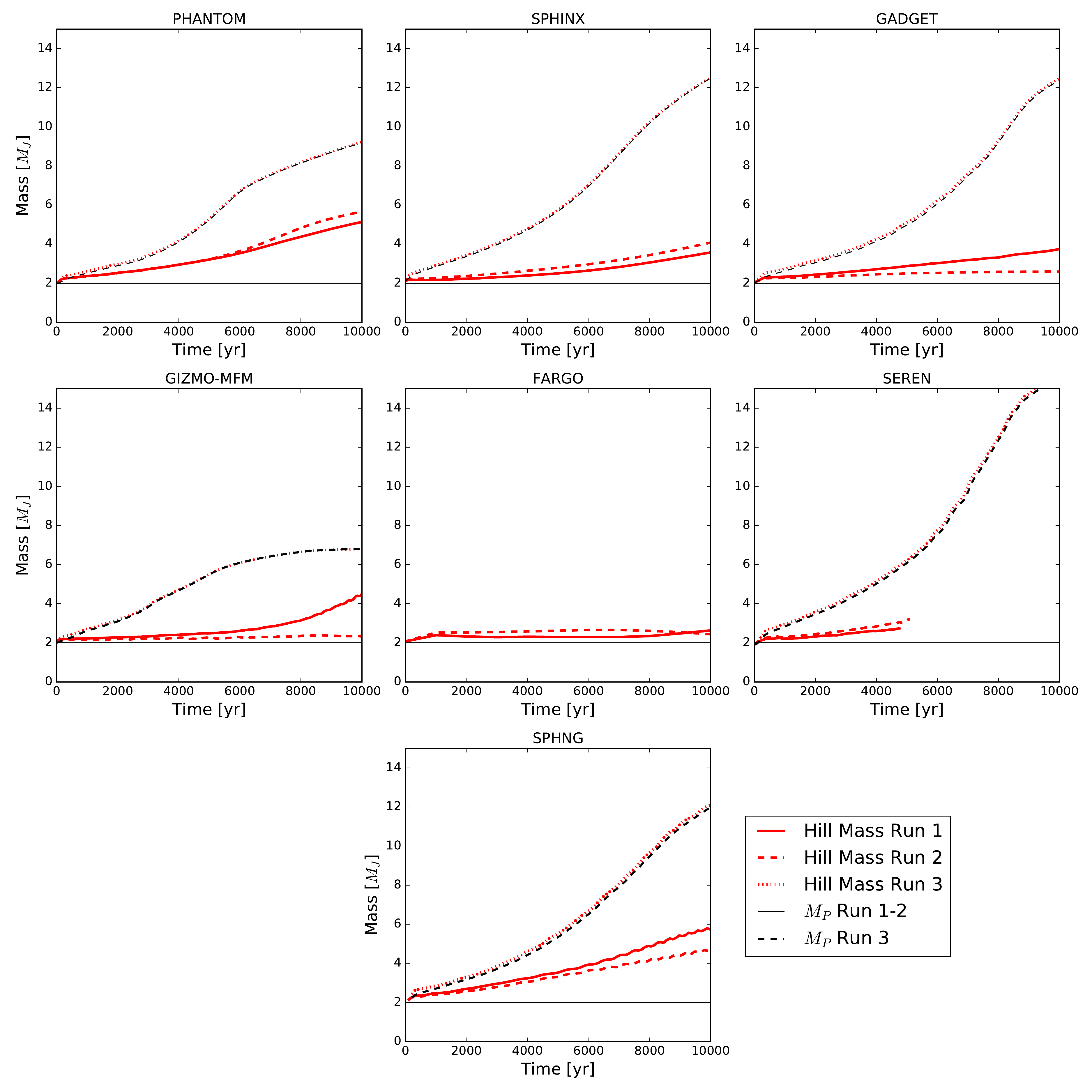}
\caption{The Hill mass, $M_{\rm H}$ (red curves), and the sink mass $M_{\rm p}$ (black), for Runs 1-3.  }
\label{fig:Rh_run13}
\end{figure*}

Fig. \ref{fig:Rh_run13} shows the Hill mass and the sink mass for Runs 1-3 calculated for the different codes. For Run 3, where gas accretion onto the sink is allowed, we see that for all the codes $M_{\rm H} \approx M_{\rm p}$. In other words, the gas mass within the Hill sphere is negligible compared with the sink mass. As the sink mass grows rapidly by gas accretion, this also means that once gas enters the Hill sphere it accretes onto the sink rapidly, so there is never a dynamically significant gas envelope around the sink. This is expected since we use a relatively large value of $R_{\rm a} = 0.5$~AU for Run 3. \cite{Nayakshin17a} found that the accretion rate onto the sink is roughly proportional to the sink radius (see Appendix in that paper) and that sink radii larger than $\sim 0.1$~AU over-estimate the rate of gas accretion onto the sink when compared with a simulation in which the clump was directly resolved\footnote{However, it is not clear what is the appropriate value of $R_{\rm a}$ to use in general as it also depends on the numerical resolution, e.g., the number of SPH particles used. Using too low a value of $R_{\rm a}$ may lead to an under-estimate of the accretion rate as the sink region may become unresolved due to a finite SPH particle resolution.}.

Fig. \ref{fig:Rh_run13} shows that in Runs 1 \& 2 the mass of gas surrounding the sink particle within $R_{\rm H}/2$ is comparable to the sink mass  by the end of the runs, in stark contrast to Run 3. For PHANTOM in particular, at $t=10,000$~yr, the Hill mass is  dominated by the envelope.

In a qualitative agreement between the codes, $M_{\rm H}$ is always larger in Run 3 than in Runs 1 and 2. This demonstrates that the gas envelope around the planet particle, which builds up in Runs 1 and 2 but not in Run 3, has a detrimental effect on further gas accretion onto the planet. This is likely due to the extra pressure of the envelope, which makes it more difficult for the gas entering the Hill sphere to remain there. However, the exact trend going from Run 1 to Run 2 in the Hill mass is not the same for the different codes. While for GADGET  and GIZMO-MFM a larger gravitational softening results in a lower mass gas envelope, this is not the case for PHANTOM and SPHINX. Therefore, gas accretion onto the planet (or the planet envelope) remains a significant source of uncertainty even in the simulations where gas accretion is turned off. An exception to this could be problems where gas accretion onto the planet is physically insignificant, such as when the planet mass is very sub-Jovian or the gas cooling time is very long \citep[as in the $\beta\gg 1$ regime in][]{Nayakshin17a}.

Let us now compare the uncertainties in the planet accretion rate versus that in migration. The left panel of fig. \ref{fig:run34} shows that there is more disagreement in the planet mass versus time plot between the different codes for Run 3 than in the planet migration tracks. The mass of gas accreted by the planet varies from a minimum of $\sim 4\mj$ to a maximum of $\sim 12\mj$, whereas the planet migration timescales vary by less than a factor of 2. We believe that this smaller disagreement in planet migration rates may be somewhat fortuitous. As the planet mass increases, the analytic formulae in the linear type I regime (e.g., eq. \ref{eq:Tanaka}) predict that the migration rate should increase linearly with planet mass. However, as the planet starts to open a gap, it starts to transition into a slower type II regime. The migration rate therefore depends on the planet mass somewhat less strongly than can be expected based on the theoretical type I predictions.

\subsection{Run 3 and Run 4}\label{sec:run34}

\subsubsection{Gap opening}\label{sec:gap}

Runs 3 and 4 both use the sink particle prescription but differ in the initial sink mass, $2\mj$ and $12\mj$, respectively. These two simulations cover the parameter space in which a growing planet goes from migrating in type I (no gap in the disc) to type II (a deep gap opened). In the outer massive disc, both planet migration rates and gas accretion rates onto the planet are far larger in the Type I regime than in the Type II regime \citep[e.g.,][]{ZhuEtal12b,Nayakshin17a}. The time and radial location where the switch between migration regimes occurs is thus of a significant importance.

\begin{figure*}
\includegraphics[width=0.99\textwidth]{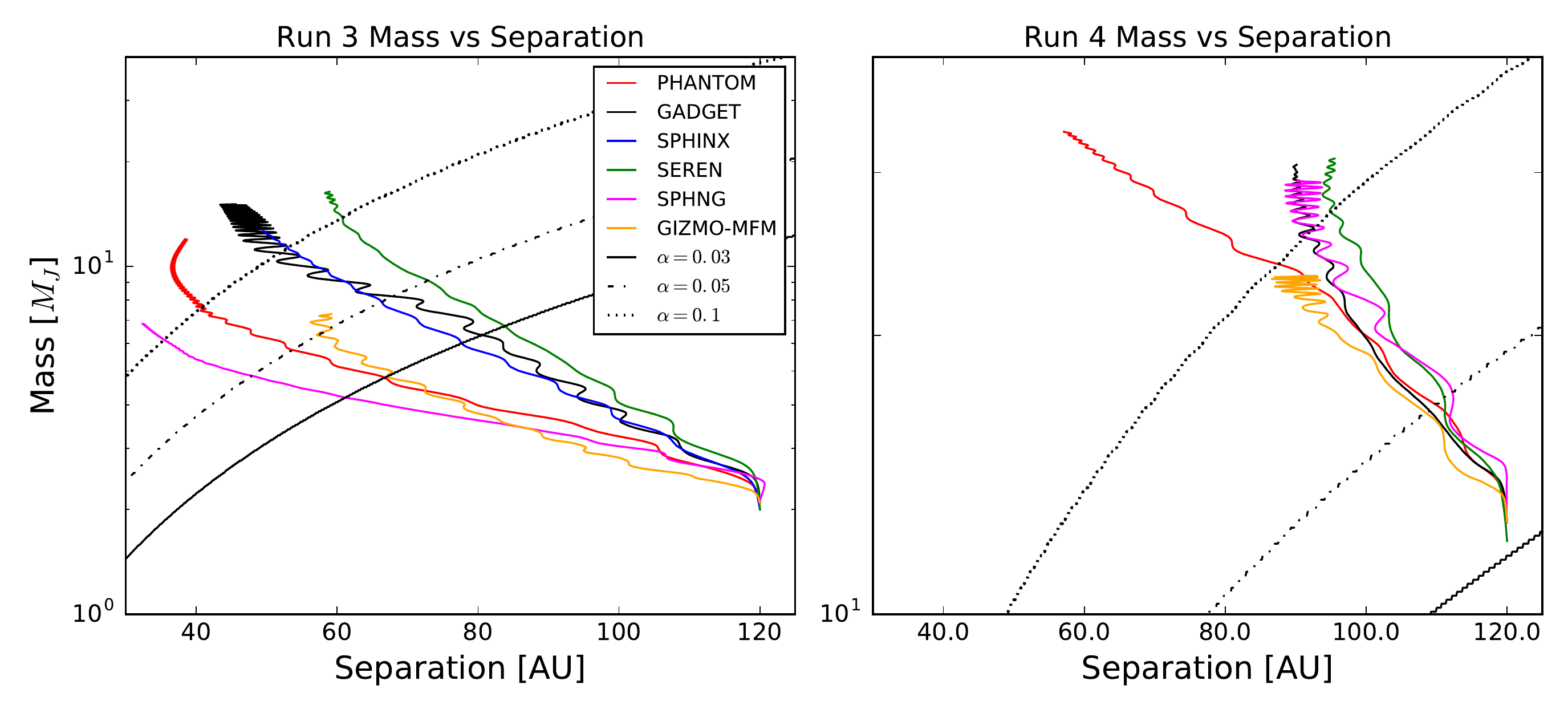}
\caption{Planet mass vs separation for Runs 3 and 4 (coloured curves). The black curves running from the bottom left to the top right corners of the panels show the gap opening planet mass (eq. \ref{eq:Crida_gap}) for several different values of the viscosity parameter $\alpha$ as specified in the legend. The planet mass-separation tracks turn more vertical when they switch into the Type II regime. As discussed in \S \ref{sec:gap}, the expected gap opening masses are given by the solid curve, but the actual ones are closer to the $\alpha=0.1$ curve.}
\label{fig:gap_opening}
\end{figure*}

Fig. \ref{fig:gap_opening} shows with different coloured lines the planet mass versus separation tracks for Runs 3 (left panel) and 4 (right panel) for all the eligible codes. The planets start at the lower right corner and move towards the upper left corner in this diagram. 

There are also four black curves in the figure that show theoretical predictions from \cite{CridaEtal06} for when a deep gap in the disc should be opened. According to these predictions, the planet opens a gap when the parameter $C_{\rm p}$ is smaller than unity:
\begin{equation}
C_{\rm p} = \frac{3}{4}\frac{H}{R_{\rm H}} + \frac{50 \alpha H^2}{R^2} \frac{M_*}{M_{\rm p}} \le 1\;.
\label{eq:Crida_gap}
\end{equation}
Here $\alpha$ is the physical viscosity parameter of the gas disc \citep{Shakura73}. We do not set a physical viscosity parameter in the runs presented here (PHANTOM offers a facility for this but most other SPH codes do not). However, artificial viscosity in numerical schemes can mimic certain effects of a physical viscosity. \cite{PriceEtal17} show that for the PHANTOM viscosity implementation, artificial viscosity parameter $\alpha_{\rm v}$, set to unity for all SPH codes here (but see \S \ref{sec:av}), results in effective \cite{Shakura73} viscosity parameter 
\begin{equation}
    \alpha = \frac{1}{10} \alpha_{\rm v} \frac{h_{\rm sml}}{H}\;,
    \label{eq:alpa_ss}
\end{equation}
where $h_{\rm sml}$ is the SPH smoothing lengh and $H$ is the local disc vertical height scale \citep[see][]{Murray96}. At the separation where our planets open gaps, we have $h_{\rm sml}/H \approx 0.4$, and hence the effective disc viscosity of these codes is about $\alpha = 0.03$. 

Additionally, self-gravitating protoplanetary discs generate physical viscosity that saturates at a maximum value of $\alpha \sim 0.06$ \citep{Gammie01,Rice05} for marginally stable discs. The value of the $Q$-parameter for our disc is significantly greater than the critical $\sim 1.5$ and we thus expect that the effective $\alpha$ from the disc self-gravity is much smaller than the maximum value.

Fig. \ref{fig:gap_opening} show the planet gap-opening mass as a function of separation for our initial discs, defined as the planet mass for which $C_{\rm p} = 1$. The solid curve sets $\alpha = 0.03$, whereas for the dashed and the dotted curves $\alpha = 0.05$ and $\alpha = 0.1$, respectively. Since planet migration effectively stalls (at least on the time scales of our simulations) when the planet switches to the type II migration regime, the radial location of this switch can be identified in the figure as the point where the planet track turns from being mainly horizontal to being more vertical. For Run 3, the left panel of fig. \ref{fig:gap_opening} shows that the location at which the migration type switches is approximately { consistent} with the \cite{CridaEtal06} prediction for $\alpha = 0.1$, although the actual value of the separation and planet mass at that point are somewhat different for the codes. However, the estimated effective disc viscosity for the codes is $\alpha  = 0.03$, and the respective (solid) curve in fig. \ref{fig:gap_opening} yields significantly smaller masses. The only exception to that is GIZMO-MFM whose meshless finite mass scheme was shown to provide smaller artificial viscosity \citep{DengEtal17}.

The results of Run 4 are largely consistent with this picture. We see that the gap opening value of planet mass and separation lie close to the $\alpha=0.1$ theoretical curve, with GIZMO-MFM transiting into type II migration somewhat earlier once again. One exception to this is PHANTOM, for which the planet seems to cross the migration type dividing line rather uneventfully.

The fact that our simulated gas clumps open gaps at higher masses and later in time than predicted by the \cite{CridaEtal06} analysis confirms the findings of \cite{MalikEtal15} who showed that in massive circumstellar discs, gap opening is more difficult than for less massive discs. As shown by \cite{MalikEtal15}, if planets migrates through the horse-shoe region faster than the gap can be excavated by planet toques, the gap remains closed even if $C_{\rm p}$ falls below unity.

Finally, although our code migration comparison project is not designed to study the longer term planet evolution that occurs in the Type II regime, we can see from fig. \ref{fig:gap_opening} that there is a significant disagreement in the planet evolution once it crosses over into the Type II regime. While qualitatively we see that planets tend to stall in Type II, as expected, some codes predict that the planets continue to migrate in while others (PHANTOM in the left hand panel) start to migrate outward. This may indicate that the secular evolution of the planets in the Type II migration regime is even more model dependent than the Type II which we mainly aim to study here.

\subsubsection{Gas accretion time scales}\label{sec:accr_times}

As emphasized by previous authors, there is a competition between the process of gas accretion onto the planet and its inward migration \citep[e.g.,][]{ZhuEtal12b,Nayakshin17a}. This competition plays a significant role in shaping of the outcome of disc fragmentation. It is hence convenient to define, in addition to the migration time scale, an accretion time scale for the planet, $t_{\rm acc}$, 
\begin{equation}
t_{acc} = \frac{M_{\rm p}}{\dot{M}} \;,
\label{eq:tacc}
\end{equation}
where $\dot M$ is the gas accretion rate onto the planet. The corresponding dimensionless quantity $\tau_{\rm acc}$,
\begin{equation}
\tau_{\rm acc} = \frac{t_{\rm acc}}{T_{\rm p}}\;,
\label{eq:tacc0}
\end{equation}
where  $T_{\rm p}= 2\pi/\Omega_{\rm p}$ is the orbital period at the planet location, will be useful as well.

\cite{BateEtal03} studied planet migration and accretion in isothermal discs and found that the following equation describes the gas accretion rate onto the planet well in the Type I migration regime,
\begin{equation}
\dot{M}_{\rm acc} = b \frac{M_{p}}{M_{*}} \Omega_{\rm p} \rho R^{3}\;,
\label{eq:Bate03a}
\end{equation}
where $b \approx 2.3$ empirically and $\rho$ is the disc midplane density. By writing $\rho = \Sigma/(2H)$ and expressing 
\begin{equation}
\Sigma = \frac{c_{\rm s}\Omega_{\rm p}}{\pi G Q}\;,
\label{eq:sigma0}
\end{equation}
where $Q$ is the Toomre parameter at the planet location, we can re-arrange the \cite{BateEtal03} result as
\begin{equation}
\tau_{\rm acc} = b^{-1} Q\;.
\label{eq:Bate_acc}
\end{equation}

\cite{ZhuEtal12a} used a 2D code to study clump migration and accretion, and provided a 2D estimate for the rate of gas accretion onto the planet, 
\begin{equation}
\dot{M} = 4 \Sigma \Omega R_{H}^{2}
\end{equation}
Expressing $\Sigma$ through eq. \ref{eq:sigma0} again, we obtain the corresponding gas accretion time scale
\begin{equation}
\tau_{acc} = \frac{M_{\rm p}}{\dot{M}T_{\rm p}} = \frac{3}{8} \frac{R_{\rm H}}{H} Q\;.
\label{eq:Zhu_acc}
\end{equation}
Since for our planets $R_{\rm H} \sim H$ within a factor of two or so, eq. \ref{eq:Zhu_acc} is actually not very different from eq. \ref{eq:Bate_acc}.

\begin{figure*}
\includegraphics[width=0.99\textwidth]{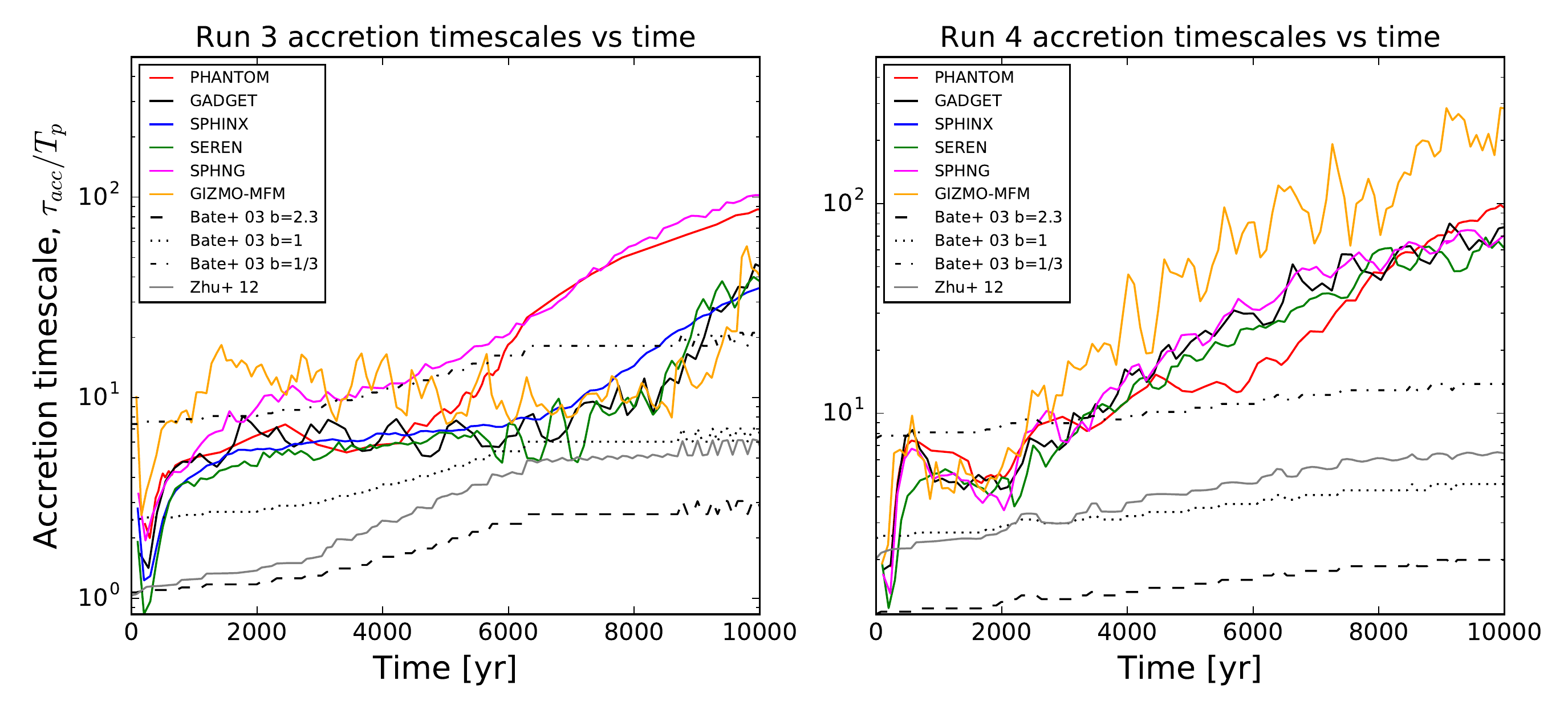}
\caption{ {\bf Left panel:} Dimensionless accretion time scale against time for Run 3. The black curves are analytical estimates for the accretion time scale given by eqs. \ref{eq:Bate_acc} (for different values of the parameter $b$) and \ref{eq:Zhu_acc}, as indicated in the legend. These estimates assume an isothermal equation of state and therefore over-predict the gas accretion rates measured in the simulations. {\bf Right panel:} Same but for Run 4. Note that the gas accretion time increases strongly when a gap in the disc is opened.}
\label{fig:acc_time}
\end{figure*}

Fig. \ref{fig:acc_time} shows dimensionless accretion time scales for Runs 3 and 4. The black  curves show the analytic estimates obtained with eqs. \ref{eq:Bate_acc} and \ref{eq:Zhu_acc}, respectively. For eq. \ref{eq:Bate_acc}, we show three curves which use $b = 2.3$ \citep[as in][]{BateEtal03}, and then also $b=$ 1, and $1/3$. We can see that both analytic prescriptions predict much faster accretion rates onto the planet than actually measured in the simulations. This is most likely due to the analytic estimates assuming an isothermal equation of state and therefore the maximum efficiency for gas capture onto the planet. In the runs presented here, the gas is not isothermal and heats up due to adiabatic compression in the Hill sphere. The cooling rate $\beta$-parameter is $\beta=10$, which is relatively large. \cite{Nayakshin17a}, see also \cite{HN18}, found that gas accretion onto planets is significantly suppressed for $\beta \gtrsim$ a few. The isothermal gas accretion rate estimates from \cite{BateEtal03} and \cite{ZhuEtal12b} physically corresponds to the $\beta \ll 1$ regime investigated in \cite{Nayakshin17a}, for which much higher accretion rates were indeed obtained. It appears that $b\approx 1/3$ in eq. \ref{eq:Bate_acc} fits the gas accretion rates in the Type I migration regime best.

Fig. \ref{fig:acc_time} also demonstrates that the accretion time increases strongly when the planet switches to the type II migration regime. This has also been seen in previous simulations \citep[e.g.,][]{BateEtal03} and is to be expected as the planet clears its immediate neighbourhood of gas, chocking its own growth. 

The initial dips in the accretion time for both panels in fig. \ref{fig:acc_time} are caused by our artificial initial conditions, in which a massive planet is injected in the disc. The gas within the Hill sphere of the planet then finds itself strongly bound to it and accretes onto the planet on a time scale shorter than the local dynamical time, $1/\Omega$. This initial transient is followed by a more self-consistent evolution in which the gas in the Hill sphere of the planet "knows about its existence".

\subsection{Importance of artificial viscosity prescription}\label{sec:av}

Artificial viscosity is used in SPH and grid based codes to treat flow discontinuities such as shocks \citep{Monaghan92,BodenheimerBook}. 
The codes we test here differ in their implementation of the artificial viscosity. Some part of the differences in the results of Runs 1-4 (discussed in \S \ref{sec:results}) may be due to these numerical technique differences. Varying the viscosity prescriptions for all of the codes would make the presentation of this paper overly long. Instead we pick one code, PHANTOM, and investigate how different artificial viscosity choices affect the results for just Run 3.

All modern SPH codes employ artificial viscosity prescriptions that include a term linear in $\Delta v$, the velocity difference between two interacting SPH particles, and a term quadratic in $\Delta v$ \citep{Springel05,PriceEtal17}. That is, the first term enters artificial viscosity with a dimensionless coefficient $\alpha_{\rm v}$, and the second with coefficient $\beta_{\rm v}$. In some codes, e.g., GADGET, these coefficients are fixed whereas in others such as PHANTOM they are allowed to vary in time during simulations. \cite{CullenDehnen10} in particular presented a method in which $\alpha_{\rm v}$ depends on the time derivative of the particle velocity divergence. The latter is used as a shock indicator and helps to eliminate artificial viscosity away from shocks, reducing unwanted numerical dissipation in dynamically quiet regions. Additionally, there are different suggestions on the appropriate values for the coefficient $\beta_{\rm v}$ to use, and in fact this may depend on the problem studied \citep{PriceEtal17}.

\begin{figure}
\includegraphics[width=0.99\columnwidth]{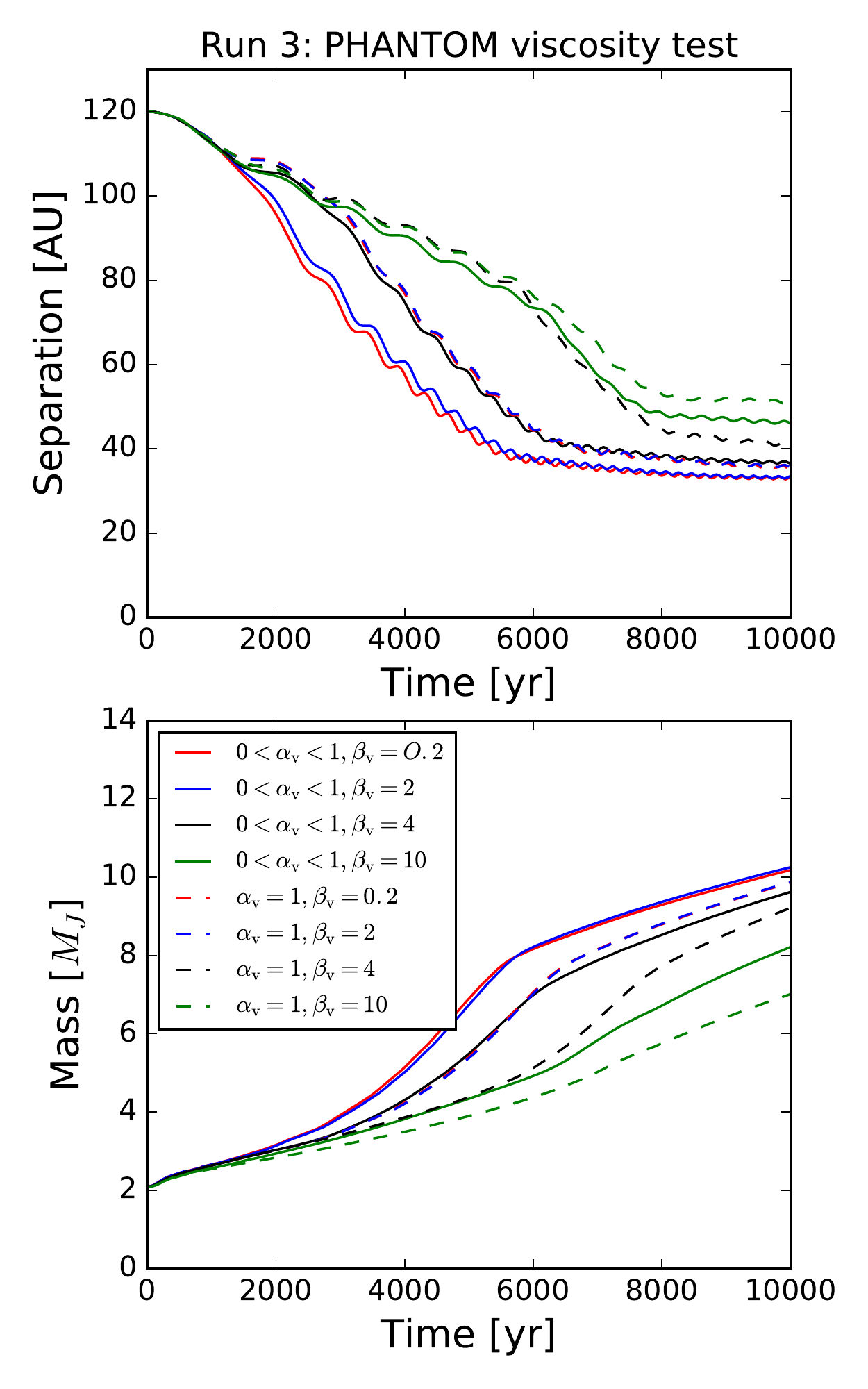}
\caption{Differences in the results of Run 3 for PHANTOM when the values of the artificial viscosity parameters $\alpha_{\rm v}$ and $\beta_{\rm v}$ are varied. See \S \ref{sec:av} for detail.}
\label{fig:visc_phantom_M2}
\end{figure}

Fig. \ref{fig:visc_phantom_M2} shows how the planet separation (top panel) and planet mass (bottom panel) are affected by the changes in the viscosity prescription for Run 3. The solid curves show Run 3 in which the $\alpha_{\rm v}$ parameter is time-dependent as in the method of \cite{CullenDehnen10}, and is allowed to vary between $0 \le \alpha_{\rm v} \le 1$. The different colours in the solid curves indicate different values of the coefficient $\beta_{\rm v}$, which we varied in a broad range, from $\beta_{\rm v} = 0.2$ to $\beta_{\rm v}=  10$. The dashed curves in fig. \ref{fig:visc_phantom_M2} show simulations with the same range in $\beta_{\rm v}$ but which now use a fixed value for $\alpha_{\rm v}=1$. 

First, without reference to the different artificial viscosity values in the figure, we note that the larger the planet mass, the more rapidly the planet migrates, at least until it opens a gap and switches to type II migration. Such a trend simply reflects the fact that more massive planets migrate more rapidly in the Type I regime (eq. \ref{eq:Tanaka}). 

Another trend obvious through all of the curves is that the higher artificial viscosity simulations tend to yield smaller gas accretion rates onto the planet. The least viscous run (red solid curve) shows the the largest gas accretion rate onto the sink and the most rapid migration. The most viscous run (green dashed curve) shows the slowest migration and the smallest gas accretion rate. The rest of the runs show a continuous transition between these two extremes.

This gas accretion trend with artificial viscosity is most likely due to the artificial viscosity heating of the gas inside the Hill radius. The larger the gas viscosity, the larger the dissipation rate within the Hill sphere, making the gas hotter. Such sensitivity of gas accretion rate onto the planet to heating within the Hill sphere was seen in the previous literature although for different reasons. \cite{NayakshinCha13} and \cite{Stamatellos15} included planet radiative feedback on the surrounding gas, and found that when the feedback is present, it keeps the gas hotter in the planet's Hill sphere, stifling gas accretion onto it. \cite{Nayakshin17a} found that slower radiative cooling rates within the Hill sphere, which also makes the gas hotter in that region, likewise leads to a reduction in the gas accretion rate.

In greater detail, we see that the runs with $\beta_{\rm v} = 0.2$ and $\beta_{\rm v} = 2$ are virtually indistinguishable, implying that the quadratic term in the artificial viscosity prescription is negligible for these small values of $\beta_{\rm v}$ for the given problem. Higher values of $\beta_{\rm v}$ however definitely affect the results. We also see that the fixed $\alpha_{\rm v}$ simulations lead to less massive and less rapidly migrating planets that tend to open a gap sooner. 

The range of migration rates and planet masses in fig. \ref{fig:visc_phantom_M2} is large enough to conclude that { although the artificial viscosity is not the only reason for differences in the results from the four runs explored in this paper, it is one of the major reasons for these differences}.  For example, GADGET's planet separation versus time track for Run 3 is similar to the green dashed curve in fig. \ref{fig:visc_phantom_M2} for PHANTOM obtained with a fixed $\alpha_{\rm v} = 1$, as used by GADGET. However, by default GADGET uses $\beta_{\rm v} = 2 \alpha_{\rm v}$, which is much smaller than $\beta_{\rm v} = 10$ for the green dashed curve. Clearly, other code differences, both in viscosity implementation \citep[GADGET uses the ][switch; PHANTOM does not]{Balsara95}, and in how artificial softening and gas accretion onto the sink is implemented must be at play. A recent study by \cite{StamatellosInutsuka18} found that the artificial viscosity coefficient $\alpha_{\rm v}$ can also drive differences in planet accretion/migration.

On the other hand, while PHANTOM simulations suggest a higher artificial viscosity might suppress accretion via spurious heating of the gas surrounding a sink particle, the trend shown by the GIZMO-MFM results in this paper suggest the role of numerical viscosity might be more complex. Indeed,
as shown in \cite{DengEtal17}, the MFM method, which does not employ
any artificial viscosity, at variance with all SPH methods, minimizes
spurious transport of angular momentum inside self-gravitating disks
and results in a lower accretion onto sink particles \citep[see Appendix B in][]{DengEtal17}. Indeed MFM solves the fluid equations  via Riemann
solver as in Godunov-type finite volume methods, which removes the
need of an artificial viscosity term in the hydro equations \citep{Hopkins15}.

Artificial viscosity implementations in SPH can induce enhanced  angular momentum transport, and thus accretion, in non-shocking rotating flows inside fluid disks, owing to the contribution of the  linear in $\Delta v$ term \citep[even with correction terms such as the Balsara switch, e.g.,][]{KaufmannEtal07}. Spurious heating and artificial angular  momentum
transport are thus two different unwanted effects of artificial viscosity which affect accretion in opposite ways. Quantifying the interplay of these two effects warrants further investigation. Nevertheless, it is noteworthy that, in the GIZMO-MFM runs, the reduced accretion limits asymptotically the mass growth of the protoplanet to less than $10 M_J$, namely within the gas giant planet regime.

\section{Comparison to population synthesis}\label{sec:fits}

At the time of writing, there are three detailed population synthesis models that address the evolution of clumps formed by gravitational instability at distances of tens to 100 AU. Such population synthesis is a necessary step to correctly interpret the results of large observational surveys \citep[e.g.,][]{ViganEtal17} { with respect to how often disc fragmentation might result in the formation of massive planets and/or brown dwarfs.}

The population synthesis models differ in assumptions about the initial state of the disc and the clumps, disc dissipation, clump radiative cooling, dust dynamics and core formation, clump migration and accretion. It is of course not possible for us to examine these different approaches here. However, we can investigate a more limited but better defined question: how well would these models reproduce the evolution of the clumps that we see in our numerical models {\em given the same disc and clump properties} as our simulations? 

To facilitate the population synthesis comparison to the simulations presented in this paper, we shall utilize the fact that the disc surface density profiles evolve relatively weakly in Run 3 as the planet remains in the Type I migration regime for most of the codes until it stalls not very far from the disc inner edge. We can therefore use the initial disc surface density profile for this comparison. For Run 4, there is a stronger surface density evolution, but we shall use the same approach (since two of the three population synthesis codes make such an approximation too), hoping that it will capture the essentials of the problem.

We first overview the clump migration approaches. \cite{ForganRice13b} use the simplified migration scheme from \cite{Nayakshin10c}, in which the Type I migration timescale is
\begin{equation}
    t_{\rm I }  =  \left( \frac{M_{\rm p}}{M_{*}} \Omega \right) ^{-1} \frac{H}{R}\;.
\end{equation}
This is derived from the \cite{Tanaka02} formula (eq. \ref{eq:Tanaka}) by requiring additionally a marginally unstable self-gravitating disc for which the Toomre parameter $Q\approx 1$ everywhere. For type II, the migration time scale is given by the disc viscous time,
\begin{equation}
     t_{\rm II}  = t_{\rm visc} = \frac{1}{\alpha \Omega} \left( \frac{H}{R} \right) ^{-2}\;.
\end{equation}
The switch between Type I and Type II migration occurs when $M_{\rm p} > M_{\rm t}$, where $M_{\rm t}$ is the transition mass given by, 
\begin{equation}
    M_{\rm t} = 2 M_{*} \left( \frac{H}{R} \right) ^{3}
\end{equation}
as used by \cite{BateEtal03}. We note that \cite{ForganEtal18} have recently presented an updated population synthesis model. We do not include this study in our code comparison here because its migration module is similar to the \cite{MullerEtal18} treatment, which is discussed below. Furthermore, \cite{ForganEtal18} also consider multiple gas clumps and model their N-body interactions. These effects can be very important in modifying the outcome of disc fragmentation \citep{HallCEtal17} but is beyond the scope of our one-clump study.

\cite{NayakshinFletcher15} use the \cite{Tanaka02} expression for type I migration written as
\begin{equation}
    t_{\rm I } = f_{\rm mig} \frac{M_{*}^{2}}{M_{\rm p}M_{\rm d}} \frac{H^{2}}{R^{2}} \Omega^{-1}
\end{equation}
where $M_{\rm d} = 2\pi \Sigma(R) R^2$ is a measure of the local disc mass, and $f_{\rm mig}$ is a dimensionless factor, set between 0.5 and 2 for different models. The factor is introduced to mimic the stochastic kicks from spiral density waves or other clumps. The Type II migration time is also set to the viscous time but with a correction multiplicative factor,
\begin{equation}
    t_{\rm II} = t_{\rm visc} \left( 1 + \frac{M_{\rm p}}{M_{\rm d}}\right)\;.
    \label{eq:FR13tII}
\end{equation}
The factor $(1+ M_{\rm p}/M_{\rm d})$ takes into account planet inertia when the disc is less massive than the planet \citep{SyerClarke95}. The correction is not very important for outer massive discs but may become large in the inner disc ($R \lesssim 10$~AU). The Crida parameter ($C_{\rm p}$, eq.~\ref{eq:Crida_gap}) is used to model the transition between Type I and Type II migration. To prevent a sharp transition when $C_{\rm p} = 1$, an exponential function of the form, $f = {\rm min}(1,\exp[-(C_{\rm p} -1)])$ is used to smooth the transition out. Note that $C_{\rm p}$ is a function of the viscosity parameter $\alpha$, which is poorly known for protoplanetary discs. \cite{NayakshinFletcher15} assumed that $\log\alpha$ is a random uniform variable in the limits between $\log(0.005)$ and $\log(0.05)$. We shall evaluate the results for these minimum and maximum values of $\alpha$.
Finally, \cite{NayakshinFletcher15} use a time-dependent 1D viscous disc model to evolve the disc surface density and other disc properties, and to conserve the angular momentum in the  interactions between the disc and the planet, but for comparison below we shall assume the initial disc properties to be consistent with the two other models. 

\cite{MullerEtal18} use a third set of equations to control planet migration, based on \cite{BaruteauEtal11}, see eq. \ref{eq:Baruteau}.
For type II migration, eq. \ref{eq:FR13tII} is used { but without the $(1 + M_{\rm p}/M_{\rm d})$ correction, which however is unimportant for this paper as it is close to unity}. \cite{MullerEtal18} also use the Crida parameter to determine when the planet switches to the type II migration, but { consider} two additional requirements for gap opening based on the work of \cite{MalikEtal15}. They define three timescales, $\tau_{\rm visc} = R^{2}/\nu$, $\tau_{\rm cross} = 2.5 R_{\rm H}v_{r}^{-1}$, where $v_r$ is the radial velocity of the planet, and $\tau_{\rm gap} = q^{2}(H/R)^{5}\Omega^{-1}$. The additional requirements demand that $\eta \tau_{\rm gap} < \tau_{\rm cross}$ and $\tau_{\rm visc} <  \tau_{\rm cross}$, where $\eta$ is a dimensionless factor { varied from 10 to 1000, with $\eta=100$ used as a baseline model. Here we test only the first of these two additional criteria since it was the one used for most of the models in \cite{MullerEtal18}.}

Finally, population synthesis models also differ in how they treat the gas accretion onto clumps.
Two of the population synthesis models \citep{ForganRice13b,NayakshinFletcher15} neglected gas accretion onto the clumps, assuming a fixed gas mass unless the clumps are tidally disrupted. \cite{MullerEtal18} prescribed a gas accretion rate onto the clumps based on earlier simulations of \cite{GalvagniMayer14}. Since our gas clumps accrete a significant amount of gas as they migrate, for a proper comparison with the population synthesis prescriptions we need all of them to take accretion into account. We previously found that the \cite{BateEtal03} expressions for gas accretion rates, when reduced down to account for a smaller accretion efficiency of our slowly cooling discs, yields a reasonable match to the ccretion time scales of our simulation (fig. \ref{fig:acc_time}). We therefore use eq. \ref{eq:Bate03a} with $b=1/3$ here to let the planets gain mass when investigating \cite{ForganRice13b,NayakshinFletcher15} models.

We also need to take into account the { decrease} in the accretion rate when the planet switches from type I to the type II regime, which is clearly seen in fig. \ref{fig:acc_time}. To this end we write
\begin{equation}
\dot{M}_{\rm p} = \dot{M}_{\rm acc} \left[ 1 + e^{-(C_{\rm p} - 1)/\Delta C} \right]^{-1}\;,
\label{eq:dotM_app}
\end{equation}
where $\dot{M}_{\rm acc}$ is the accretion rate estimate given by eq. \ref{eq:Bate03a} where $\Delta C=0.2$. We shall see below that this yields a decent fit to the  planet mass evolution for both Run 3 and Run 4. With this approach, the comparison of population synthesis models to hydrodynamical simulations isolates just the planet migration and gap opening aspects.

\begin{figure*}
\includegraphics[width=0.99\textwidth]{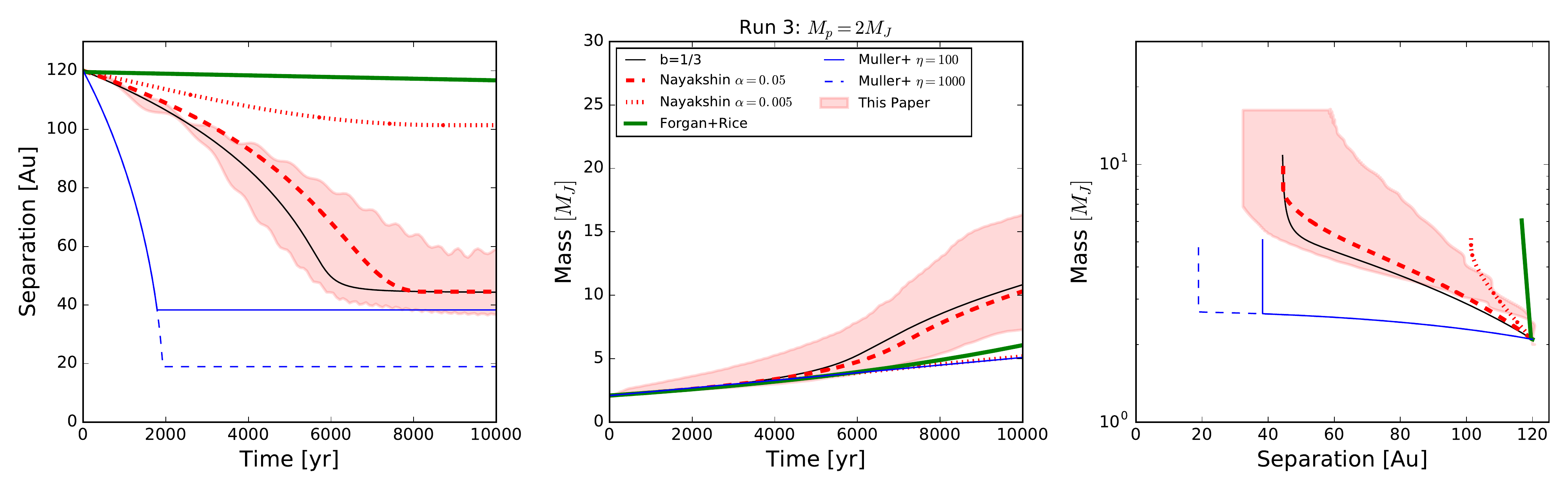}
\caption{Comparison of migration and accretion tracks for Run 3, shown as a shaded region, with population synthesis models as shown in the legend. 
{\bf Left panel:} planet separation vs time; {\bf Middle panel:} planet mass vs time; {\bf Right panel:} mass vs separation.}
\label{fig:run3_fit}
\end{figure*}

\begin{figure*}
\includegraphics[width=0.99\textwidth]{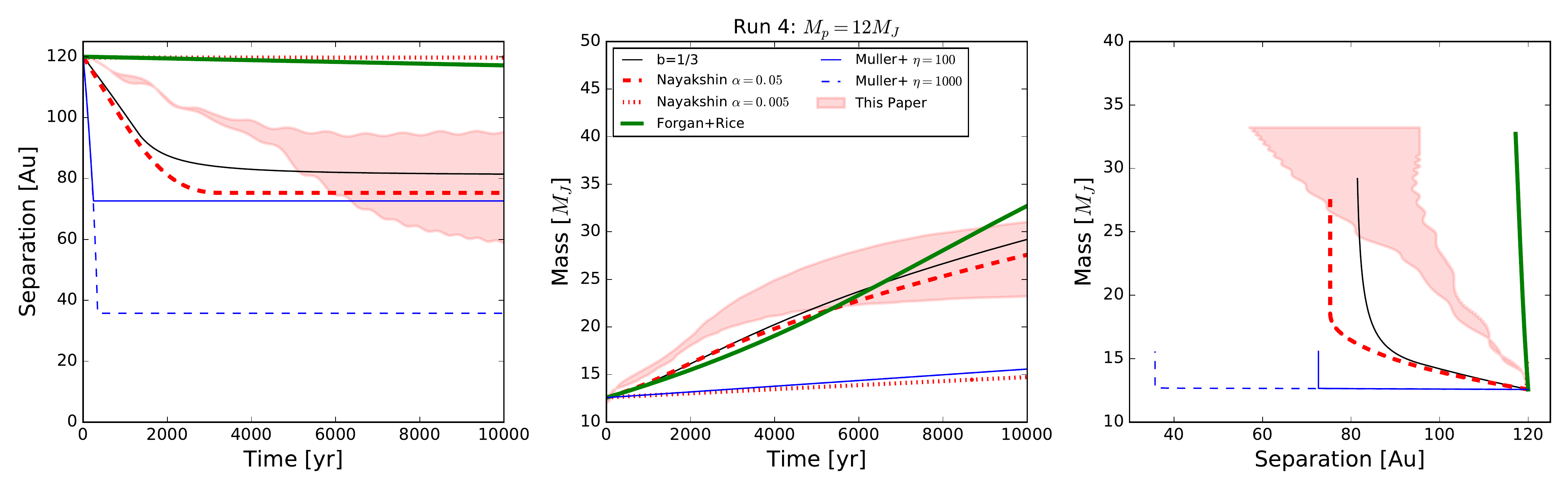}
\caption{Same as fig. \ref{fig:run3_fit} but for Run 4. See \S \ref{sec:fits} for more detail.}
\label{fig:run4_fit}
\end{figure*}

Figs. \ref{fig:run3_fit} and \ref{fig:run4_fit} show such comparisons for Run 3 and Run 4, respectively. The shaded region represents approximately the range of numerical results obtained for these runs with the different numerical codes. In particular,  PHATOM and GIZMO-MFM are selected to show the fastest and the slowest migrating planets for Run 3 in the left panel; the SEREN and GIZMO-MFM curves to show the range of models in the middle and right panels. For Run 4, PHANTOM and SEREN are selected as the extremes for the both planet accretion and migration tracks. 

We see that there is a significant difference in how the three population synthesis models compare to the numerical results. The \cite{MullerEtal18} study appears to over-estimate somewhat how quickly and how far the planets migrate before they switch into the Type II regime. This seems to be because \cite{MullerEtal18} formulae are based on \cite{BaruteauEtal11} and yield too rapid migration by a factor of a few in the type I regime, as was seen in fig. \ref{fig:migration_timescales}. Also, the planet opens a gap a little closer to the star than it does in the simulations. This depends on the parameter $\eta$ which is set to 100 and 1000 for the solid and the dashed curves, respectively.

The \cite{ForganRice13b} approach appears to yield too slow a migration rate. This is because the planet switches into the Type II migration rate immediately, as the transition mass in this approach is set to $2 (H/R)^3 M_* \approx 2\mj$, and the planet is already this massive in the beginning of the Run 3. This under-estimates the planet transition mass, which is found to be in the range of $\sim 7 \mj$  to $\sim 30\mj$ (cf. fig. \ref{fig:gap_opening}). We found that a much better fit to Run 3 is obtained with the \cite{ForganRice13b} formulae if the transition mass is increased by a factor of $\sim 5$. 

The \cite{NayakshinFletcher15} formulae used \cite{Tanaka02} expression for the migration rate with a dimensionless factor $f_{\rm migr}$ in front. The factor was a logarithmically uniform random variable in the limits $0.5 < f_{\rm migr} < 2$ and was meant to { mimic} possible stochastic kicks that the clumps obtain when interacting with the spiral density waves of the disc \citep[see][]{BaruteauEtal11}. In the interest of figure clarity we use $f_{\rm migr}=1$ in fig. \ref{fig:run3_fit} for this model. Further, \cite{NayakshinFletcher15} use the \cite{CridaEtal06} switch for gap opening, with the $\alpha$ parameter being a sum of two parts, a constant $\alpha$ and a part driven by self-gravity. We neglect the latter contribution to $\alpha$ here, and show two cases with $\alpha = 0.005$ and $0.05$ in fig. \ref{fig:run3_fit}. It is apparent that the smaller $\alpha$ curve (red solid) opens a gap in the disc far more easily than expected. The $\alpha = 0.05$ curve (red dashed) { seems more reasonable}.  However, we must remember that \cite{NayakshinFletcher15} neglected gas accretion onto the planet. The agreement of their prescriptions with Run 3 would have been worse if we kept the planet mass fixed at $2\mj$.


Fig. \ref{fig:run4_fit} shows that for a more massive gas clump none of the population synthesis prescriptions fare particularly well. The \cite{ForganRice13b} model and the low viscosity model of \cite{NayakshinFletcher15} open a gap in the disc too early, as for Run 3. The higher viscosity model of \cite{NayakshinFletcher15} does relatively well in terms of gap opening mass but over-estimates the speed with which the planet migrates in initially. The \cite{MullerEtal18} equations also yield clumps migrating in too rapidly, and the gap is opened too close in compared with numerical simulations. 

We therefore conclude that matching numerical results with analytic expressions remain a problem. What is particularly alarming is that seemingly benign changes in the parameters of the population synthesis prescriptions (such as a factor of a few change in the planet transition mass) can yield planet migration rates different by $\sim$ two orders of magnitude as the planet switches into the Type II migration regime prematurely.

%
%

\section{Discussion and Conclusions}\label{sec:discussion}

\subsection{Numerics}

In this paper we set up four different simulations of a gas planet starting at an initial separation of 120 AU in a massive gaseous disc. These 4 Runs differed in treatment of gas accretion onto the planet and the initial planet mass. We then performed these simulations with seven different numerical codes in order to compare their results.

We find differences by a factor of $\sim 2$, and sometimes as large as 3, between different codes in the accretion and migration rates. A more detailed analysis using PHANTOM indicates that these differences are to a large degree due to variations in the artificial viscosity prescriptions between the codes, although other factors such as gravitational softening and sink particle treatment probably also contribute.

We also compared our results with the planet migration and accretion prescriptions from three previous population synthesis studies (\S \ref{sec:fits} and figs. \ref{fig:run3_fit} \& \ref{fig:run4_fit}). The \cite{ForganRice13b} approach is found to open deep gaps in the disc prematurely. Since planets migrate very slowly in the type II regime, this implies that this study may over-estimate the population of gas giants remaining at wide separation after gas discs are dispersed. The \cite{MullerEtal18} study, on the other hand, over-estimates the rate of inward migration of planetary mass clumps. The \cite{NayakshinFletcher15} study fits the Run 3 results relatively well in the high viscosity case but not for the low viscosity case. In the latter case, clumps open deep gaps in the disc and tend to stall on wide orbits when they should migrate to smaller radii via Type I migration. However, all three population synthesis prescriptions fare poorly for Run 4 in which a more massive planet is considered. Additionally, \cite{ForganRice13b} and \cite{NayakshinFletcher15} neglect gas accretion onto clumps.

\subsection{Observational implications}

Recent observational surveys of solar type stars show that only a few \% of such stars are orbited by massive planets or brown dwarfs on orbits larger than $\sim 10$~AU \citep[e.g.,][]{BillerEtal13,ChauvinEtal15,ViganEtal15,ViganEtal17}. Let us call this fraction $N_{\rm present}$. This is a key constraint on the theory of planet and brown dwarf formation via gravitational instabilities of large massive gas discs. However, it is even more important to consider the frequency of such objects in a time-integrated sense, that is, the number of gas clumps formed by disc fragmentation  per star. Let this fraction be $N_{\rm birth}$. The two fractions are clearly connected via
\begin{equation}
N_{\rm present} = N_{\rm birth} \times P_{\rm surv}\;,
\end{equation}
where $P_{\rm surv} < 1$ is the probability for a gas clump to survive to the present day at a wide separation. 

Detailed calculations and population synthesis approaches are necessary to calculate $P_{\rm surv}$ accurately. \cite{ForganRice13b} obtained $P_{\rm surv} \sim 1$, \cite{NayakshinFletcher15} had $P_{\rm surv} \simlt 0.1$ \citep[][found a yet smaller value,
$P_{\rm surv}  \sim 0.05$, when { feedback effects of the luminous core} onto the clump are included]{Nayakshin16a}, and \cite{MullerEtal18} found $P_{\rm surv}\ll 1$ but noted that this depends strongly on model assumptions. \cite{RiceEtal15} in addition showed that N-body interactions with secondary stars may remove a number of wide separation planets, lowering the fraction of $P_{\rm surv}$ further in the post-disc dispersal phase.

Our simulations and population synthesis comparison (figs. \ref{fig:run3_fit} and \ref{fig:run4_fit}) demonstrate that just varying the assumptions about the underlying physics of the disc or clumps {\em by a factor of a few} may influence the results very strongly. One has to also add to this that the exact birth mass of the fragments and the mass of the disc at which it fragments are not known to better than a factor of a few \citep[e.g.,][]{KratterL16}, and the evolution of the clump strongly depends on uncertain disc cooling and dust physics \citep{Nayakshin17a}, radiative feedback from the clump \citep{NayakshinCha13,Stamatellos15,MercerStamatellos17}, etc. Therefore, the uncertainty in $P_{\rm surv}$ at present is uncomfortably large. At this time { we cannot rule out a survival probability that would imply $N_{\rm birth} > 1$.}

What is the best way forward in resolving these uncertainties? Clearly, theoretical and simulation efforts to constrain $P_{\rm surv}$ from first principles should continue. However, other indirect approaches can also help. If the migration processes allow GI planets to populate the whole range of separations between the stellar radius and  their birth place, what would be the differences between the objects left behind from this migration and those made by Core Accretion? If we are able to understand these differences more robustly, then discovering (or not) such unusual objects at separations less than 10 AU may yield independent constraints on $P_{\rm surv}$ and $N_{\rm birth}$.

%
%


\section*{Acknowledgements}

Theoretical astrophysics research at the University of Leicester is supported
by a STFC grant. The work performed at the University of Leicester used the ALICE High Performance Computing Facility, and the DiRAC Data Intensive service at Leicester, operated by the University of Leicester IT Services, which forms part of the STFC DiRAC HPC Facility (www.dirac.ac.uk). The equipment was funded by BEIS capital funding via STFC capital grants ST/K000373/1 and ST/R002363/1 and STFC DiRAC Operations grant ST/R001014/1. DiRAC is part of the National e-Infrastructure.

FM acknowledges support from The Leverhulme Trust, the Isaac Newton Trust and the Royal Society Dorothy Hodgkin Fellowship.  This work was undertaken on the COSMOS Shared Memory system at DAMTP, University of Cambridge operated on behalf of the STFC DiRAC HPC Facility. This equipment is funded by BIS National E-infrastructure capital grant ST/J005673/1 and STFC grants ST/H008586/1, ST/K00333X/1.  This work also used the DiRAC Data Centric system at Durham University, operated by the Institute for Computational Cosmology on behalf of the STFC DiRAC HPC Facility (www.dirac.ac.uk). This equipment was funded by a BIS National E-infrastructure capital grant ST/K00042X/1, STFC capital grant ST/K00087X/1, DiRAC Operations grant ST/K003267/1 and Durham University. DiRAC is part of the National E-Infrastructure.

\bibliographystyle{mnras}
\bibliography{nayakshin}



\bsp	
\label{lastpage}
\end{document}